\documentclass[a4paper]{IEEEtran}

\usepackage[utf8]{inputenc} 
\usepackage[T1]{fontenc}
\usepackage{url}
\usepackage{ifthen}
\usepackage{cite}
\usepackage[cmex10]{amsmath} 
\usepackage{amssymb}
\usepackage{amsfonts}
\usepackage{cleveref}
\usepackage{dsfont}
\usepackage{notation}
\usepackage{verbatim,booktabs}
\usepackage{graphicx}
\newtheorem{theorem}{Theorem}
\newtheorem{theorem*}{Theorem*}
\newtheorem{corollary}{Corollary}
\newtheorem{lemma}{Lemma}

\newtheorem{definition}{Definition}
\usepackage[dvipsnames]{xcolor}

\DeclareMathOperator*{\argmax}{arg\,max}
\DeclareMathOperator*{\argmin}{arg\,min}

\IEEEoverridecommandlockouts

\begin{document}

\title{A Sphere-Packing Error Exponent\\ for Mismatched Decoding} 

\author{Ehsan Asadi Kangarshahi and Albert Guill\'en i F\`abregas
		\thanks{E. Asadi Kangarshahi is with the Department of Engineering, 
			University
			of Cambridge, Cambridge CB2 1PZ, U.K. (e-mail: ea460@cam.ac.uk).
			
			A.~Guill\'en i F\`abregas is with the  Department of Engineering, 
			University
			of Cambridge, Cambridge CB2 1PZ, U.K. and the Department of Information and 
			Communication
			Technologies, Universitat Pompeu Fabra, Barcelona 08018, Spain (e-mail: guillen@ieee.org). 
			
			This work was supported in part by the European Research Council under 
			Grant 725411 and by the Spanish Ministry of Economy and Competitiveness under Grant PID2020-116683GB-C22.
		}
		\thanks{This work has been presented in part at the 2021 IEEE International Symposium on Information Theory, Melbourne, Australia.
		}
	}

\maketitle

\begin{abstract}
We derive a sphere-packing error exponent for coded transmission over discrete memoryless channels with a fixed decoding metric. By studying the error probability of the code over an auxiliary channel, we find a lower bound to the probability of error of mismatched decoding. The bound is shown to decay exponentially for coding rates smaller than a new upper bound to the mismatch capacity which is established in this paper. For rates higher than the new upper bound, the error probability is shown to be bounded away from zero. The new upper bound is shown to improve over previous upper bounds to the mismatch capacity. 
\end{abstract}

\section{Introduction}

Communication problems where the receiver needs to employ a suboptimal decoder are typically cast within the mismatched decoding framework \cite{scarlett2020information}. These situations arise when optimal maximum-likelihood decoding cannot be used: i) the channel transition is unknown and imperfectly estimated or, ii) when, for complexity reasons, the channel likelihood is too difficult to compute and an alternative decoding metric is needed. In addition, some important problems in information theory like the zero-error or zero-undetected error capacities can be cast as instances of mismatched decoding \cite{csiszarDMC}. In the mismatched decoding problem, the optimal maximum-likelihood decoder is replaced by a maximum metric decoder, in which the metric is not necessarily the channel likelihood. For a fixed channel $W$ and decoding metric $\metric$, finding a single-letter expression for the mismatch capacity $C_{\metric}(W)$ remains an open problem and only bounds are known.

 A number of single-letter lower bounds have been derived in the literature \cite{huimis,csiszarDMC,csiszargraph,Merhav} (see also \cite{scarlett2020information} for a recent survey). A number of lower bounds based on multiuser coding techniques have been derived \cite{lapidoth,somekh2014achievable,scarlett2016multiuser}, some yielding improvements over standard single-user coding. Most of these lower bounds have been derived via random coding which in turn yield single-letter lower bounds on the error exponent.
As suggested by \cite{csiszarDMC}, multiletter versions of achievable rates can yield strict improvements over their single-letter counterparts.

Instead, up until recently, not much progress had been made on upper bounds. Balakirsky \cite{balakirsky1995converse} claimed that for binary-input discrete memoryless channels (DMC), the mismatch capacity coincided with the lower bound in \cite{huimis,csiszargraph}. Reference \cite{counterexample} provided a binary-input ternary-output counterexample to this converse invalidating its claim. In particular, it was shown that the order-$2$ multiletter version of the multiuser coding rate in \cite{somekh2014achievable,scarlett2016multiuser} is strictly higher than the bound derived by in \cite{huimis,csiszargraph}. In \cite{EsiAlbert} (see also \cite{EsiAlbertISIT2019}), we proposed a single-letter upper bound to the mismatch capacity based on transforming the channel in such a way that errors on the auxiliary channel imply a mismatched-decoding error in the original channel. Reference \cite{somekh2022single} cast the bound in \cite{EsiAlbert} as multicast transmission over a broadcast channel, significantly simplifying the proof. The bounds in \cite{somekh2022single} improved over that in \cite{EsiAlbert} in several directions. In addition, \cite{somekh2022single} also provided conditions that a pair of channel and decoding metric must fulfil for the bound to be tight and thus give the mismatch capacity. Recently, further improvements were presented in \cite{anelia_isit2021,isit_2021_long}. Reference \cite{anelia_isit2021} builds on the idea of multicast transmission allowing the possibility that when an error is made in the auxiliary channel, a mismatched decoding error in the original channel is made with a certain probability, instead of deterministically as in \cite{EsiAlbert,somekh2022single}. The bound in \cite{isit_2021_long} is a preliminary part of this work and also relaxes this condition in a different way and will be discussed in detail in this paper. All bounds  \cite{EsiAlbert,somekh2022single,anelia_isit2021,isit_2021_long} belong to the same family of constrained minimizations of the mutual information of an auxiliary channel, and can be expressed as 
\begin{align}
C_{\metric}(W) \leq \max_{P_X}\min_{\substack{P_{Y\hat{Y}|X} \in \Mc\\ P_{Y|X} = W}} I(P_X,P_{\hat{Y}|X}),
\end{align}
where the set $\Mc$ quantifies the statistical relationships among the channel input $X$, output $Y$ and auxiliary channel output $\hat Y$, ensuring, either deterministically or probabilistically, that errors in the auxiliary channel induce mismatched decoding errors in the true channel. The set $\Mc$ may depend on the input distribution $P_X$. Therefore, it is of interest to enlarge the set $\Mc$ of joint conditional distributions $P_{Y\hat{Y}|X}$, or broadcast channels, such that the aforementioned error condition is fulfilled.

Not many single-letter upper bounds on the error exponent of mismatched decoding are available, other than the trivial upper-bounds to the standard channel coding problem. In a recent paper \cite{bondaschi2021mismatched}, the authors proved that the expurgated error exponent derived in \cite{scarlett2014expurgated} at rate zero is tight for a wide class of channels and decoding metrics. 
In this paper, we derive a sphere-packing upper bound to the error exponent of mismatched decoding. We also show that the rate where the sphere-packing upperbound becomes equal to zero is a new upper bound on the mismatch capacity.

This paper is organized as follows. Section \ref{sec:prelim} introduces the notation and preliminary concepts. Section \ref{sec:sp} introduces the main results of the paper, the new upper bound to the mismatch capacity, the sphere-packing bound to the error exponent and a comparison of the new bound and previously proposed bounds. Section \ref{sec:opt} discusses an optimization interpretation of the design of the set $\Mc$. Proofs of the main results can be found in Section \ref{mainproofs}. Proofs of auxiliary results can be found in the appendices.

\section{Preliminaries}
\label{sec:prelim}

We consider reliable communication over a DMC $W$ defined over input and output alphabets $\Xc = \{1,2,\dotsc,J\}$ and $\Yc = \{1,2,\dotsc,K\}$. We denote the channel transition probability by $W(k|j)$. A codebook $\Cc_n$ is defined as a set of $M$ sequences $\Cc_n = \big\{\xv_1,\dotsc,\xv_M\big\}$, where $\xv_m= \big(x_{1,m},\dotsc,x_{n,m}\big)\in\Xc^n$, for $m \in \{1,\dotsc,M\}$. 
A message $m \in \{1,\dotsc,M\}$ is  chosen equiprobably and $\xv_m$ is sent over the channel. The channel produces a noisy observation $\yv=(y_1,\dotsc,y_n)\in\Yc^n$ according to $W^n(\yv|\xv) = \prod_{i = 1}^{n}W(y_i|x_i)$. 

Upon observing $\yv\in\Yc^n$ the decoder produces an estimate of the transmitted message $\hat m \in \{1,\dotsc,M\}$. The average and maximal error probabilities are respectively defined as 
\begin{align}
	&P_e(\Cc_n) = \PP[\hat m\neq m] \\
	&P_{e,\rm max}(\Cc_n) = \max_{m\in \{1,\dotsc,M\}}\PP[\hat m\neq m|m \text{ is sent}].
\end{align}
The decoder that minimizes the error probability is the maximum-likelihood (ML) decoder, that produces the message estimate $\hat m$ according to
\beq
\hat m = \argmax_{m\in\{1,\dotsc,M\}} W^n\big(\yv|\xv_m\big).
\eeq
Rate $R>0$ is said to be achievable if for any $\epsilon > 0$ there exists a sequence of length-$n$ codebooks $\{\Cc_n\}_{n = 1}^{\infty}$ such that $|\Cc_n| \geq 2^{n(R-\epsilon)}$, and $ \liminf_{n \to \infty}P_e(\Cc_n)= 0$. The capacity of $W$, denoted by $C(W)$, is defined as the largest achievable rate.

In situations with channel uncertainty, it is not possible to use ML decoding and instead, the decoder produces the message estimate $\hat m$ as
\beq
\hat m = \argmax_{m\in\{1,\dotsc,M\}} \metricn\big(\xv_m,\yv\big),
\eeq
where $\metricn\big(\xv,\yv\big) =\sum_{i=1}^n q\big(x_i, y_i\big)$ and $q:\Xc\times\Yc\to\RR$ is the decoding metric. We  refer to this decoder as $\metric$-decoder. When $\metric(x,y)=\log W(y|x)$, the decoder is ML, otherwise, the decoder is said to be mismatched \cite{huimis,csiszarDMC,csiszargraph,Merhav,scarlett2020information}. 
The average and maximal error probabilities of codebook $\Cc_n$ under $\metric$-decoding are respectively denoted by $P_e^\metric(\Cc_n,W)$ and  $P_{e,\rm max}^\metric(\Cc_n,W)$.  The mismatch capacity $C_\metric(W)$  is defined as supremum of all achievable rates with $\metric$-decoding.

The method of types \cite[Ch. 2]{csiszar2011information} will be used extensively in this paper. We recall some of the basic definitions and introduce some notation. The type of a sequence $\xv = (x_1,x_2,\dotsc,x_n) \in \Xc^n$ is a column vector representing its empirical distribution, \textit{i.e.}, $ \hat{\pv}_{\xv} (j) = \frac{1}{n} \sum_{i = 1}^{n}\indicator\{x_i = j\}$. The set of all types of $\Xc^n$ is denoted by $\Pc_n(\Xc)$. For $\pv_X \in \Pc_n(\Xc)$, the type class $\Tc(\pv_X)$ is set of all sequences in $\Xc^n$ with type $\pv_X$, $\Tc(\pv_X) = \{\xv \in \Xc^n\,|\, \hat{\pv}_{\xv} = \pv_X\}$. 
The joint type of sequences $\xv = (x_1,x_2,\dotsc,x_n)\in \Xc^n$ and $\yv=(y_1,y_2,\dotsc,y_n)  \in \Yc^n$ is defined as a matrix representing their empirical distribution 
\begin{align}
	\hat{\pv}_{\xv\yv}(j,k) = \frac{1}{n}\sum_{i = 1}^{n}\indicator\{x_i = j,y_i=k\}.
\end{align} 
The set of joint types on $\Xc^n,\Yc^n$ is given by $\Pc_n(\Xc\Yc)$.
The conditional type of $\yv$ given $\xv$ is the matrix
\begin{align}
\hat{\pv}_{\yv|\xv}(k|j) = \begin{cases} 
\frac{\hat{\pv}_{\xv\yv}(j,k)}{\hat{\pv}_{\xv}(j)} \
 &\hat{\pv}_{\xv}(j)> 0\\
\frac{1}{|\Yc|} \ &\text{otherwise.}
\end{cases}
\end{align}
The set of conditional types on $\Yc^n$ given $\Xc^n$ is denoted by $\Pc_n(\Yc|\Xc)$. For $\pv_{Y|X} \in \Pc_n(\Yc|\Xc)$ and sequence $\xv \in \Tc(\pv_X)$, the conditional type class $\Tc_{\xv}(\pv_{Y|X})$ is defined as $\Tc_{\xv}(\pv_{Y|X}) = \{\yv \in \Yc^n\,|\, \hat{\pv}_{\yv|\xv} = \pv_{Y|X}\}.$

Similarly, we can define the joint type of $\xv,\yv,\hat{\yv}$, as the empirical distribution of the triplet. For $j \in \Xc$ and $k_1,k_2 \in \Yc$,
 \begin{align}
 \hat{\pv}_{\xv\yv\hat{\yv}}(j,k_1,k_2) = \frac{1}{n}\sum_{i = 1}^{n} \indicator\{x_i = j,y_i = k_1,\hat{y}_i = k_2\}.
 \end{align}
We define the joint conditional type of $\yv,\hat{\yv}$ given $\xv\in\Tc(\pv_X)$ as
 \begin{align}\label{definetype}
 \hat{\pv}_{\yv\hat{\yv}|\xv}(k_1,k_2|j) = \begin{cases} 
\frac{\hat{\pv}_{\xv\yv\hat{\yv}}(j,k_1,k_2)}{\hat{\pv}_{\xv}(j)} \
 &\hat{\pv}_{\xv}(j)> 0\\
\frac{1}{|\Yc|}\indicator\{k_1=k_2\} \ &\text{otherwise.}
\end{cases}
\end{align}
The set of all joint conditional types is denoted by $\Pc_n(\Yc\hat{\Yc}|\Xc)$. Additionally, for $\pv_{Y\hat{Y}|X} \in \Pc_n(\Yc\hat{\Yc}|\Xc)$ we define: 
 \begin{align}\label{lfnrnfgrn}
 \Tc_{\yv\xv}(\pv_{Y\hat{Y}|X}) = \{\hat{\yv} \in \Yc^n \,|\, \hat{\pv}_{\yv\hat{\yv}|\xv} = \pv_{Y\hat{Y}|X}\}.
 \end{align}
Throughout the paper use the notation $\pv_{Y}\pv_{X|Y}$ to denote the distribution $P_{XY}$ defined by
\begin{align} \label{dafeiafmafgrngs}
P_{XY}(j,k) = \pv_{Y}(k)\pv_{X|Y}(j|k)
\end{align}
Note that the former multiplication for two generic types is not necessarily a type, therefore we denote the result with the probability distribution notation.

The mutual information and conditional relative entropy are respectively defined as
\begin{align}
I(P_X,P_{Y|X}) &\triangleq \EE \Big[\log\frac{P_{Y|X}(Y|X)}{\sum_{x'} P_X(x') P_{Y|X}(Y|x')}\Big],\\
D(P_{Y'|X}\|P_{Y|X}|P_X) &\triangleq \sum_{x \in \Xc} P_X(x)  \cdot D(P_{Y'|X = x}\|P_{Y|X=x}).
\end{align}
\begin{definition}
	A random variable $X$ is called sub-Gaussian with parameter (sub-Gaussian norm) $\theta$ if for any $\xi >0$ we have
	\begin{align}
		\PP[|X - \EE[X]| \geq \xi] \leq \Gamma e^{ \frac{-\theta^2\xi^2}{2}},
	\end{align}
where $\Gamma$ is a constant. Throughout the paper we use $\Gamma = 2$ for simplicity of notation. Additionally, $\Gamma=2$ is sufficient for the relevant random variables to our proofs. Therefore, we use the following definition instead of the previous one
\begin{align}
	\PP[|X - \EE[X]| \geq \xi] \leq 2e^{ \frac{-\theta^2\xi^2}{2}}.
\end{align} 
\end{definition}

\begin{definition}
	Let $P,Q$ be probability distributions defined on alphabet $\Xc$. Then, the infinity norm between $P,Q$ is defined as
	\begin{align}
		|P-Q|_{\infty} = \max_{x \in \Xc} |P(x) - Q(x)|.
	\end{align}
\end{definition}
Throughout the paper and proofs, whenever we use $|P-Q|_{\infty}$ we will implicitly assume that $P$ is absolutely continuous with respect to $Q$ and vice versa.

\begin{definition}
	Let $\Cc_n = \{\xv_1,\xv_2,\dotsc,\xv_M \}$ be a codebook and $\pv_{Y|X}$ be a conditional type. The maximum type-conflict error probability is defined as
	\begin{align}
		&\petcemax(\Cc_n, \pv_{Y|X}) \notag\\
		&~~~~\eqdef \max_{m\in\{1,\dotsc,M\}} \PP\Big[ \bigcup_{{\bar m} \neq m} \{ \hat{\pv}_{\yv|\xv_m} = \hat{\pv}_{\yv|\xv_{\bar m}} = \pv_{Y|X}  \}\Big],
	\end{align}
	where the probability is with respect to the uniform distribution over the type class $\Tc_{\xv_m}(\pv_{Y|X})$.
\end{definition}

Similarly to \cite{EsiAlbert}, the main idea of this paper is to relate the type-conflict error performance of a given codebook over an auxiliary channel $V$ with the $\metric$-decoding performance of the same code over channel $W$. The main reason for studying type-conflict errors is that an equation of the form $\hat \pv_{\yv|\xv_2} =\hat \pv_{\yv|\xv_1}$ provides more information about the properties of the error than ML errors, where we simply have a scalar likelihood inequality. In addition, it can be shown that for rates $R>C(V)$, then the probability of type-conflict errors bounded away from zero.

We proceed by introducing a few definitions. Recall the definition of maximal set from \cite{EsiAlbert}. Consider the set 
\beq \label{vbaofdewmd}
\Sc_\metric(k_1,k_2)\eqdef\big \{ j\in \Xc | j=\argmax_{j'\in\Xc}\metric(j',k_2) - \metric(j',k_1)\big\}.  
\eeq
A  joint conditional distribution $P_{Y\hat{Y}|X}$ is said to be maximal if for all $(j,k_1,k_2) \in \Xc \times \Yc \times \Yc$,
  \begin{align} \label{grre}
	P_{Y\hat{Y}|X}(k_1,k_2|j) = 0 ~\text{ if } ~j \notin \Sc_\metric(k_1,k_2).
  \end{align}
The set of all maximal joint conditional distributions was defined to be $\Mmax(\metric)$. In this work, for a given distribution $P_{X_1}$, we define the set of maximal joint conditional distributions as follows.

\begin{definition}\label{cafekggo44}
	 $\Mmax(\metric,P_{X_1})$ is the set of all joint conditional distributions $P_{Y\hat{Y}|X_1}$ such that
	\begin{align}
		\min_{\substack{P_{X_2|X_1\hat Y}: \\ X_2-X_1\hat Y-Y \\P_{\hat{Y}X_2} = P_{\hat{Y}X_1}}} \EE [\metric(X_2,Y)] \geq \EE [\metric(X_1,Y)],
		\label{eq:maximaldef}
	\end{align}
	where the notation $X_2-X_1\hat Y- Y$ denotes that $X_2, (X_1\hat Y)$ and $Y$ form a Markov chain. 
\end{definition}
We close this section by showing that that $\Mmax (\metric) \subset \Mmax(\metric,P_{X_1})$ for any input distribution $P_{X_1}$. Assume that $P_{Y\hat Y |X_1} \in \Mmax (\metric)$. Then from \cite[Lemma 3]{EsiAlbert} we have for any $X_2$ such that $P_{\hat Y X_1} = P_{\hat Y X_2}$ 
\begin{align}
	\EE[\metric(X_2, Y)] \geq \EE[\metric(X_1,Y)].
\end{align}
This implies that $P_{Y\hat Y |X_1}$ satisfies \eqref{eq:maximaldef} and as a result, $P_{Y\hat Y |X_1} \in \Mmax (\metric,P_{X_1})$. 
As we will show, this enlarged set of maximal distributions yields an improved upper bound on the mismatch capacity. Throughout the paper, we have taken the convention that $X_1$ represents the sent codeword and $X_2$ represents an auxiliary codeword.

\section{Main Results}
\label{sec:sp}
In this section, we introduce over the main results of this paper. We first introduce an upper bound to the mismatch capacity.
\begin{theorem} \label{th:main2}
	Let $W,q$ be channel and decoding metric, respectively. Then, 
	\begin{align} 
		C_{\metric}(W) \leq \bar R(W,\metric).
		\label{eq:upper_bound}
	\end{align}
	where 
	\begin{align} \label{eq:cqbound}
		\bar R(W,\metric) \eqdef \max_{P_{X}} \min_{\substack{P_{Y\hat{Y}|X} \in \Mmax(\metric,P_{X})\\ P_{Y|X} = W}} I(P_{X},P_{\hat{Y}|X}),
	\end{align}
\end{theorem}
\begin{IEEEproof}
See section \ref{fepfepdkwea} for the proof of this theorem.
\end{IEEEproof}

\begin{corollary} \label{cnaefneuf}
	If some joint conditional distribution $P_{Y\hat Y|X} \in  \Mmax(\metric,P_{X})$ for all input distributions $P_X$, then
	\begin{align}
		C_\metric(W) \leq C( P_{\hat Y | X}).
	\end{align}
\end{corollary}
The next result introduces a sphere-packing upper bound to the error exponent of mismatched decoding.
\begin{theorem} \label{dafbeghr}
	Consider a fixed composition codebook $\Cc_n$ with length $n$, rate $R$ and composition $\pv_X$. The error probability of $\Cc_n$  with $\metric$-decoding over channel $W$ satisfies
	\begin{align}
		-\frac{1}{n}\log \pemax(\Cc_n,W) \leq \Espq(\pv_X,R+\zeta_n) - \delta_{n},
		\label{eq:spb}
	\end{align}
	where 
	\begin{align} \label{vnsefiaefieafn}
		\hspace{-2mm}\Espq(P_X,R) = \min_{\substack{P_{Y'\hat Y|X} \in \Mmax(\metric,P_{X}) \\  I(P_X,P_{\hat{Y}|X}) \leq R }} D(P_{Y'|X}\|P_{Y|X}|P_X)
	\end{align}
	and
	\begin{align}
		\zeta_{n} &= (JK-1) \frac{\log(n+1)}{n} +\frac{\log2}{n}\label{eq:zetan}\\
		\delta_{n} &= \Oc \Big( \frac{\log n}{n} \Big)\label{eq:deltan}
	\end{align}
\end{theorem}
\begin{IEEEproof}
See section  \ref{defoefpesss} for the proof of this theorem.
\end{IEEEproof}

	Next we introduce the analogous version of Theorem \ref{dafbeghr} for a family of type dependent metrics. Firstly we define the analogous version of $\Mmax$ for type dependent metrics. With a slight abuse of notation we use $\metric(\pv_{XY})$ to denote a type-dependent metric $\metric$ computed for type $\pv_{XY}$.

\begin{definition}
	Set $\Mmax^{\rm td}(\metric,P_{X_1})$ is defined as follows
	\begin{align}
		&\Mmax^{\rm td}(\metric,P_{X_1}) \notag\\
		&~~~~\eqdef \Bigg \{ P_{Y\hat Y|X_1} \bigg| \min_{\substack{P_{X_2|X_1 \hat Y}: \\ X_2- \hat Y X_1 - Y \\ P_{\hat Y X_1} = P_{\hat Y X_2} }}  \metric(P_{X_2Y}) \geq \metric(P_{X_1Y}) \Bigg \}
	\end{align}
\end{definition}

Consider type-dependant metrics $\metric(P_{XY})$ where $\metric$ is convex in $P_{Y|X}$ when $P_X$ is fixed. This is an important family since important metrics such as maximum mutual information (MMI) metric defined as $\metric(P_{XY}) = I(P_{XY})$ have this property. For this family of metrics, we have exactly the same statement as that of Theorem \ref{dafbeghr}, but replacing $\Mmax(\metric,P_{X_1})$ by $\Mmax^{\rm td}(\metric,P_{X_1})$. 
See Section \ref{sec:tdm} for the proof of this statement.

\subsection{Example} \label{grofeodeod}
In this part we show the application of our bound to the counterexample in \cite{counterexample}, where the channel and metric are  
\begin{align}
	W &= \begin{bmatrix}
		0.97 &0.03  &0\\
		0.1  &0.1   &0.8
	\end{bmatrix}\\
	q &= \begin{bmatrix}
		0  &0   &0\\
		0  &\log(0.5) &\log(1.36)
	\end{bmatrix}.
\end{align}
For this example $C(W) = 0.7133$ bits/use, the rate achievable by $2$-letter superposition coding from \cite{counterexample} is $R_{\rm sc}^{(2)}(W,q)=0.1991$ bits/use and our previous converse \cite{EsiAlbert} stated that $C_q(W)\leq \bar R_\metric(W)= 0.6182$ bits/use. 
By numerically solving the optimization problem in \eqref{eq:maximaldef} we observe the  joint conditional distribution given in Table \ref{tbl:maximal_example2} is maximal for all input distributions $P_X$.
\begin{table}[h]
	\centering
	\caption{Nonzero entries of $P_{Y\hat{Y}|X}$ for Example 1}
	\label{tbl:maximal_example2}
	\begin{tabular}{llll}
		\toprule
		$(j,k_1,k_2)$ & $P_{Y\hat{Y}|X}$ & $(j,k_1,k_2)$ & $P_{Y\hat{Y}|X}$ \\
		\midrule
		$(1,1,1)$ & $0.3778$ & $(2,1,1)$ & $0.1000$ \\
		$(1,1,2)$ & $0.5922$ & $(2,2,2)$ & $0.0911$ \\
		$(1,2,2)$ & $0.0300$ & $(2,3,3)$ & $0.6956$ \\
		& & $(2,3,2)$ & $0.1133$\\
		\bottomrule
	\end{tabular}
\end{table}

Marginalizing the above $P_{Y\hat{Y}|X}$ over $Y$ we obtain
\begin{align}
	P_{\hat{Y}|X} = \begin{bmatrix}
		0.3756 &0.6244  &0\\
		0.1  &0.2044   &0.6956
	\end{bmatrix}.
\end{align}	
Therefore, by using Corollary \ref{cnaefneuf} we have
\begin{align}
	C_\metric(W) &\leq C( P_{\hat Y | X}) \\
	&= 0.4999 \text{ bits/use}.
\end{align}

Observe that the above result can be further improved by solving the optimization problem in \eqref{eq:cqbound}.
In terms of computation, unlike the bound proposed in \cite{EsiAlbert}, optimizing \eqref{eq:cqbound} is not a simple task. This observation stems from the fact that the maximal set $\Mmax(\metric,P_X)$ in  \eqref{eq:cqbound} depends on $P_X$, unlike the maximal set $\Mmax(\metric)$ in \cite{EsiAlbert}. In addition, the set $\Mmax(\metric,P_X)$ is itself defined as an optimization problem over distributions $P_{X_2|X\hat Y}$ and this makes the problem more difficult than  \cite{EsiAlbert}. As illustrated next, the advantages of the bound in \eqref{eq:cqbound} are potentially significant even under the conditions of Corollary \ref{cnaefneuf}.

\subsection{Comparison with other bounds} \label{sec:comparison}
In this section, we compare the new bound to the mismatch capacity given in \eqref{eq:cqbound}  with some of the recent bounds that have appeared in the recent literature. Recall that all bounds have the same form
\begin{align}
	C_{\metric}(W) \leq \max_{ P_X}\min_{\substack{P_{Y\hat{Y}|X} \in \Mc\\ P_{Y|X} = W}} I(P_X,P_{\hat{Y}|X}),
\end{align}
where $\Mc$ is a set of joint conditional distributions. In the following, we compare the set 
\begin{align}
	&\Mmax(\metric,P_{X_1}) \notag\\
	&~~~\eqdef \Bigg \{ P_{Y\hat Y|X_1}\,:\, \min_{\substack{P_{X_2|X_1 \hat Y}: \\ X_2- \hat Y X_1 - Y \\ P_{\hat Y X_1} = P_{\hat Y X_2} }}  \EE[\metric(X_2,Y)] \geq \EE[ \metric(X_1,Y)] \Bigg \}
	\label{eq:setdef}
\end{align}
defined in Definition \ref{cafekggo44} with those from previously proposed bounds. In particular, we show that all previously proposed bounds are such that $\Mc \subset \Mc_{\rm max}(\metric,P_X)$.

To begin with, we compare our recent upper bound derived in \cite{EsiAlbert}. The expression of the set $\Mc$ is
\begin{align}
	&\Mc_{\rm max}(\metric)\notag\\
	&~= \big\{P_{Y\hat Y|X_1} \,:\,  P_{Y\hat Y|X_1} (k_1, k_2|j)=0 ~\text{if}~ j\notin\Sc_q(k_1,k_2)\big\},
\end{align}
where
\beq
\Sc_q(k_1,k_2) = \big \{ j\in \Xc | j=\argmax_{j'\in\Xc}\metric(j',k_2) - \metric(j',k_1)\big\}.
\eeq
From \cite[Lemma 3]{EsiAlbert}, we know that if $P_{Y\hat{Y}|X_1} \in \Mc_{\rm max}(\metric)$ then
\begin{align} \label{cbeifeofemf}
	\forall  X_1: P_{\hat Y X_1} = P_{\hat Y  X_2} \Rightarrow\EE[ \metric( X_2,Y)] \geq \EE[\metric(X_1,Y)].
\end{align}
However, $P_{Y\hat{Y}|X_1} \in \Mc_{\rm max}(\metric)$ is not a necessary condition for \eqref{cbeifeofemf} to hold.
Instead, for $P_{Y\hat{Y}|X_1} \in \Mc_{\rm max}(\metric,P_{X_1})$ we have 
\begin{align} \label{vafnefienf}
	\forall  X_2:  X_2-X_1\hat Y&-Y,  P_{\hat Y X_1}= P_{\hat Y  X_2} \notag\\
	&\Rightarrow\EE[ \metric( X_2,Y)]\geq \EE[\metric(X_1,Y)].
\end{align}
We observe that $P_{Y\hat{Y}|X_1} \in \Mc_{\rm max}(\metric,P_{X_1})$  is both a necessary and sufficient condition for \eqref{vafnefienf} being true. Therefore, we see thanks to the additional Markov chain constraint on $\tilde X$,  $\Mc_{\rm max}(\metric)  \subset \Mc_{\rm max}(\metric, P_{X_1})$. Indeed, the more constraints are added to \eqref{eq:setdef}, the more joint conditional distributions are able to satisfy the inequality, thus making the set larger.

Two improved upper bounds of the same family were presented in \cite{somekh2022single}. The first bound is expressed as
\begin{align}
	C_{\metric}(W) \leq \min_{\substack{P_{Y\hat{Y}|X} \in \Gamma(\rho, \metric)\\ P_{Y|X} = W}} C_{\rho}(P_{\hat Y|X}),
\end{align}
where
\begin{align}
	&\Gamma(\rho, \metric)= \{P_{Y\hat{Y}|X} | P_{Y\hat{Y}|X}(y,\hat y |x) = 0 ~\text{ if } ~x \notin \Sc_{\rho,\metric}(y,\hat y)\}, \\
	&\Sc_{\rho, \metric}(y,\hat y) = \big \{ x' \in \Xc | x' \notin \argmax_{x\in\Xc} \rho(x,\hat y) - \metric(x,y)\big\}.
\end{align}
The expression of the second bound, which is also valid for type-dependent metrics, is given by
\begin{align}
	C_{\metric}(W) \leq \max_{P_X} \min_{\substack{P_{Y\hat{Y}|X} \in \Gamma^*(\metric, P_{X})\\ P_{Y|X} = W}} I(P_{X},P_{\hat{Y}|X}),
\end{align}
where
\begin{align}
	\Gamma^*(\metric, P_{X}) &= \big\{P_{Y\hat Y|X} \,|\, \forall\, V_{Y\hat Y X \tilde X}: V_{Y\hat Y X} \ll P_{X} \times P_{Y\hat Y|X},\notag\\
	&V_{\hat YX} = V_{\hat Y \tilde X} \Rightarrow \EE[\metric(\tilde X,Y)]\geq\EE[\metric(X,Y)]  \big \},
\end{align}
and $V_{Y\hat Y X} \ll P_{X} \times P_{Y\hat Y|X}$ denotes $P_{X} \times P_{Y\hat Y|X}$ being absolutely continuous with respect to $V_{Y\hat Y X}$.
The second bound was shown to be stronger than the first one \cite{somekh2022single}, and we therefore focus on the comparison with the second. By expressing the set $\Mc_{\rm max}(\metric, P_{X})$ with a similar notation we get
\begin{align} \label{cvaifneifaomd}
	\Mc_{\rm max}(\metric, P_{X}) &= \big\{P_{Y\hat Y|X} | \forall P_{Y\hat Y X \tilde X}: P_{\hat YX} = P_{\hat Y \tilde X},\notag\\
	& \tilde X-X\hat Y-Y \Rightarrow \EE[\metric(\tilde X,Y)]\geq\EE[\metric(X,Y)] \big\}.
\end{align}
Observe that the constraint $\EE[\metric(\tilde X,Y)]\geq \EE[\metric(X,Y)] $ in the set $\Gamma^*(\metric, P_{X})$  should hold for all $V_{Y\hat Y X \tilde X}$ with $V_{Y\hat Y X} \ll P_{X} \times P_{Y\hat Y|X}, V_{\hat YX} = V_{\hat Y \tilde X}$. Instead, the constraint $\EE[\metric(\tilde X,Y)]\geq \EE[\metric(X,Y)] $ in the set $\Mc_{\rm max}(\metric, P_{X})$ must hold for all $P_{Y\hat Y X \tilde X}$ such that distribution of $Y\hat Y X$ is equal to $P_{Y\hat Y X}$ but $\tilde X$ is further constrained by the Markov chain property $\tilde X-X\hat Y-Y$. 
Therefore, similarly to the previous comparison, we find that 
\begin{align}			
	\Gamma^*(\metric, P_{X})  \subset \Mc_{\rm max}(\metric, P_{X}).
\end{align}

More recently, a further improvement was reported in \cite{anelia_isit2021}.
The main bound in \cite{anelia_isit2021} is expressed as 
\begin{align}
	C_{\metric}(W) \leq \max_{P_X} \min_{\substack{P_{Y\hat{Y}|X} \in \Theta^*(\metric, P_{X})\\ P_{Y|X} = W}} I(P_{X},P_{\hat{Y}|X}) 
\end{align}
where $\Theta^*$ is defined as
\begin{align}
	\Theta^*(\metric, P_{X}) = \{P_{Y\hat Y|X} &\,|\, \forall P_{Y\hat Y X \tilde X}: P_{\hat YX} = P_{\hat Y \tilde X} \notag\\
	&\Rightarrow \EE[\metric(X,Y)] \leq \EE[\metric(\tilde X,Y)] \}
\end{align}
By comparing the set $\Theta^*(\metric, P_{X})$ to $\Mc_{\rm max}(\metric, P_{X})$, we find that the constraint $\EE[\metric(\tilde X,Y)]\geq \EE[\metric(X,Y)] $ in the set $\Theta^*(\metric, P_{X})$ must hold for all $P_{Y\hat Y X \tilde X}$ such that distribution of $Y\hat Y X$ is equal to $P_{Y\hat Y X}$, but a further constraint on $\tilde X$ is missing. Since $\Mc_{\rm max}(\metric, P_{X})$ has an additional Markov chain constraint on $\tilde X$, we have that
\begin{align}
	\Theta^*(\metric, P_{X})  \subset \Mc_{\rm max}(\metric, P_{X}).
\end{align}

\section{Alternative Interpretation of Maximal Sets} \label{sec:opt}

So far, every joint conditional distribution $P_{Y\hat Y|X}$ that belongs to the corresponding maximal set 
\begin{align}
&\Mmax(\metric,P_{X_1})\notag\\
& \eqdef \Bigg \{ P_{Y\hat Y|X_1}\,:\, \min_{\substack{P_{X_2|X_1 \hat Y}: \\ X_2- \hat Y X_1 - Y \\ P_{\hat Y X_1} = P_{\hat Y X_2} }}  \EE[\metric(X_2,Y)] \geq \EE[ \metric(X_1,Y)] \Bigg \}
\label{eq:setdef2}
	\end{align}
from Definition \ref{cafekggo44} yields a valid upper bound to the mismatch capacity. The joint conditional distribution that minimizes the mutual information of the auxiliary channel yields the best bound. This is the case because maximal joint conditional distributions are such that if an error is made over the auxiliary channel $P_{\hat Y|X}$, then a mismatched decoding error is made on the original channel $P_{ Y|X}$, constrained to be $P_{ Y|X}=W$. This latter statement also holds for a significant fraction of the errors, not necessarily all. 

In this section, we discuss a different approach to the construction of the maximal set. Specifically, we first fix the auxiliary channel $V = P_{\hat Y|X}$, and then optimize the resulting joint conditional distribution to fulfill the maximality constraint, i.e., if an error is made over the auxiliary channel $V=P_{\hat Y|X}$, then a mismatched decoding error is made on the original channel $W=P_{ Y|X}$. This naturally gives maximal set of auxiliary channels. Not fixing to the joint conditional distribution between $V,W$ offers the possibility to derive a potentially stronger upper bound. Specifically,  we  first consider the type $\pv_{\hat Y X_1X_2}^\star$ from Lemma \ref{bcafkamd} such that for most type conflict errors on channel $V$, the empirical type $\hat \pv_{\hat \yv \xv_1 \xv_2}$ is equal to $\pv_{\hat Y X_1X_2}^\star$. Then given this type, we can optimize the joint conditional distribution $P_{Y\hat Y |X}$ to fulfill the maximality condition for type $\pv_{\hat Y X_1X_2}^\star$. This is in contrast to only knowing the type conflict error condition over the auxiliary channel, i.e., $\hat \pv_{\hat \yv \xv_1} = \hat \pv_{\hat \yv \xv_2}$ for every joint type  $\hat \pv_{\hat \yv \xv_1 \xv_2}$.

The above interpretation of the construction of the maximal set, suggests to define it as the following set of all auxiliary channels $V$ 
 \begin{align} \label{okgkrtarw}
 &\Vc_{\rm max}(\metric,P_{X_1}) \notag\\
 &\eqdef \Bigg \{ V\,:\,	\max_{\substack{P_{Y\hat Y |X_1}: \\ P_{\hat Y |X_1} = W \\ P_{\hat Y |X_2} = V }} \min_{\substack{P_{X_2|X_1\hat Y} :\\ X_2- \hat Y X_1 - Y \\ P_{\hat Y X_1} = P_{\hat Y X_2} }}  \EE[\metric(X_2,Y)] \geq \EE[\metric(X_1,Y)]\Bigg\},
 \end{align}
 where the inner minimization corresponds to the choice of type $\pv_{\hat Y X_1X_2}^\star$ and the outer maximization refers to the choice of the joint conditional  distribution with marginals $V,W$. Observe that $\EE[\metric(X_1,Y)]$ is constant for any given $P_{X_1}$, channel $W$ and metric $\metric$. The following lemma implies that this alternative definition gives the same bounds to the mismatch capacity and error exponent as those described in previous sections for additive decoding metrics. 
 \begin{lemma} \label{lemmagame}
 	 	The  optimization order in \eqref{okgkrtarw} can be exchanged. More precisely,
 	\begin{align}
 		&\max_{\substack{P_{Y\hat Y |X_1}:\\ P_{ Y |X_1} = W\\P_{\hat Y |X_1} = V }} \min_{\substack{X_2 \\ X_2- \hat Y X_1 - Y \\ P_{\hat Y X_1} = P_{\hat Y X_2} }}  \EE[\metric(X_2,Y)] \notag\\
		&~~~~~~~~~~~= \min_{\substack{P_{X_2|X_1\hat Y} :\ \\ X_2- \hat Y X_1 - Y \\ P_{\hat Y X_1} = P_{\hat Y X_2} }}  \max_{\substack{P_{Y\hat Y |X_1}: \\ P_{ Y |X_1} = W\\P_{\hat Y |X_1} = V }}  \EE[\metric(X_2,Y)].
 	\end{align}
 \end{lemma}

\begin{IEEEproof}
We show the following 3 facts in order to prove the lemma.
\begin{itemize}
\item
The set of $P_{Y\hat Y |X_1}$ where $P_{Y |X_1} = W,P_{\hat Y |X_2} = V$ is convex. This is evident, since marginalizing a probability distribution is a linear operation.
\item  The set of $P_{X_2|X_1\hat Y}$ such that $ X_2- \hat Y X_1 - Y$ and $P_{\hat Y X_1} = P_{\hat Y X_2}$ is a convex set. To prove this statement we fix $P_{Y\hat Y |X_1}$ and consider the set of joint probability distributions $P_{Y\hat Y X_1 X_2}$  
\begin{align} \label{vsdfjaowfjefoafw}
	\big\{P_{Y\hat Y X_1 X_2} | P_{\hat Y X_1} &= P_{\hat Y X_2},\notag\\
	& P_{Y\hat Y X_1 X_2} = P_{X_2|\hat Y X_1} P_{Y\hat Y X_1 } \big \}
\end{align} 
Therefore, if two random variables $\tilde X_2, \bar X_2$ both have joint probability distributions in the above set \eqref{vsdfjaowfjefoafw}, then, any new random variable $\hat X_2$ drawn according to $P_{\hat X_2|\hat Y X_1} = \alpha P_{\tilde X_2|\hat Y X_1} + (1-\alpha) P_{\bar X_2|\hat Y X_1}$ is in set \eqref{vsdfjaowfjefoafw}.
\item Finally, we need to show $\EE[\metric(X_2,Y)]$ is linear in both $P_{Y\hat Y |X_1}$ and $P_{ X_2|\hat Y X_1} $ when fixing either  of them.
This is proven by expanding $\EE[\metric(X_2,Y)]$
\begin{align}
\EE[\metric(X_2,Y)]&= \sum_{x_1,x_2,y, \hat y} \metric(x_2,y) P_{X_1}(x_1)\notag\\
&\times P_{Y\hat Y |X_1}(y,\hat y|x_1) P_{X_2|\hat Y X_1}(x_2|\hat y, x_1),
\end{align}

which is linear both in $P_{Y\hat Y |X_1}$ and $P_{X_2|\hat Y X_1}$ when we fix either of the two.

As a result, we have a convex-concave optimization problem, and therefore, by the minimax theorem \cite{minimax}, the order of optimization can be exchanged. 
\end{itemize} 
\end{IEEEproof}

Observe that, as a consequence of the above lemma, every joint conditional distribution $P_{Y\hat Y |X_1}\in\Mmax(\metric,P_{X_1})$, then the corresponding $P_{\hat Y|X_1}\in \Vc_{\rm max}(\metric,P_{X_1})$. Conversely, for every $P_{\hat Y|X_1}\in \Vc_{\rm max}(\metric,P_{X_1})$, there exists a joint conditional distribution $P_{Y\hat Y |X_1}\in\Mmax(\metric,P_{X_1})$. Therefore, the optimization problems involving $\Mmax(\metric,P_{X_1})$ or $\Vc_{\rm max}(\metric,P_{X_1})$ in the calculation of the upper bound to the mismatch capacity and error exponent give the same result.

We next illustrate how this argument continues to hold for the optimization of the error exponent for type-dependent metrics, but not necessarily for the upper bound to the mismatch capacity. In particular, for type-dependent metrics, consider the following set of auxiliary channels
 \begin{align} \label{okgkrtarw2}
 &\Vc_{\rm max}^{\rm td}(\metric,P_{X_1},W) \notag\\
 &~~\eqdef \Bigg \{ V\,:\,	\max_{\substack{P_{Y\hat Y |X_1}: \\ P_{\hat Y |X_1} = W \\ P_{\hat Y |X_2} = V }} \min_{\substack{P_{X_2|X_1\hat Y} :\\ X_2- \hat Y X_1 - Y \\ P_{\hat Y X_1} = P_{\hat Y X_2} }} q(P_{X_2Y}) \geq q(P_{X_1Y})\Bigg\}.
 \end{align}
We have the following result for the error exponent.
\begin{theorem}\label{th:3}
	Consider a fixed composition codebook $\Cc_n$ with length $n$, rate $R$ and composition $\pv_X$. The error probability of $\Cc_n$  with a type-dependent metric decoder using $\metric$  over channel $W$ satisfies
	\begin{align}
		-\frac{1}{n}\log \peq(\Cc_n,W) \leq \Espq(\pv_X,R+\zeta_n) - \delta_{n},
	\end{align}
	where we have
	\begin{align}
		&\Espq(P_X,R)\notag\\
		&~~~~ = \min_{P_{Y'|X}}~\min_{\substack{V \in\Vc_{\rm max}^{\rm td}(\metric,P_{X},P_{Y'|X}) \\  I(P_X,V) \leq R }} D(P_{Y'|X}\|P_{Y|X}|P_X)
	\end{align}
	and $\zeta_{n} ,\delta_{n}$ are defined in \eqref{eq:zetan}, \eqref{eq:deltan}, respectively.
\end{theorem}

\begin{IEEEproof}
See Section \ref{sec:tdm}.
\end{IEEEproof}

The rate where the the exponent becomes equal to zero is the following
\begin{align}
 \max_{P_{X}}  \min_{ \substack{V \in\Vc_{\rm max}^{\rm td}(\metric,P_{X},W)}} I(P_{X},V).
 \label{eq:lkjhglkj}
\end{align}
Unfortunately,  the analysis of Section \ref{mainproofs} for this expression fails to work. The main reason is that the error probability may in principle decay subexponentially for rates above  \eqref{eq:lkjhglkj}, and the techniques to prove the mismatch capacity upper bound of  Section \ref{mainproofs} are not sufficient. 

\section{Proofs of the Main Results} \label{mainproofs}

\subsection{Proof of Theorem 1} \label{fepfepdkwea}
We will use the following results proved in Appendix \ref{auxiliary_theorems}.

\begin{theorem} \label{maintheorem}
	Let $\Cc_n = \{\xv_1,\dotsc,\xv_M \}$ be a constant composition codebook of composition $\pv_{X}$ and length $n$. Assume that $P_{Y\hat Y |X} \in \Mmax(q, \pv_X)$ is a maximal joint conditional distribution. Then, there exists a joint conditional distribution $\bar P_{Y\hat Y|X} \in \Mmax(q, \pv_X)$ satisfying
	\begin{align}
		&\bar P_{\hat Y|X} \times \pv_X \in \Pc_n(\Xc\times\Yc) \label{eq:pbar1}\\
		&|\bar P_{\hat Y|X} \times \pv_X - P_{\hat Y|X} \times \pv_X|_{\infty} \leq \frac{1}{n}\label{eq:pbar2} \\
		& |\bar P_{Y|X} \times \pv_X - P_{Y|X} \times \pv_X|_{\infty} \leq \frac{K}{n},\label{eq:pbar3}
	\end{align}	
	and a constant $\gamma > 0$ that depends only on $P_{Y|X}$ and $\metric$ such that
	\begin{align}\label{maintheoremeq}
		\pemax(\Cc_n, \bar P_{Y|X}) \geq \gamma \petcemax(\Cc_n,\bar P_{\hat Y|X}).
	\end{align}
\end{theorem}

The next result from \cite{EsiAlbert} lower bounds the type-conflict error probability.
\begin{theorem}\label{th:3}
	Under the assumptions of Theorem \ref{maintheorem}, for every type $\pv_X$, there exist $n_0,\bar E(R)>0$ such that if $n>n_0$ and $ \frac{1}{n}\log |\Cc_n|> I(P_X,P_{\hat Y|X}) $
	\begin{align}
		\petcemax(\Cc_n,P_{\hat Y|X}) \geq 1 - 2^{-n\bar E(R)}.
	\end{align}
\end{theorem}

	We show that for any $R>\bar R(W,\metric)$ there exist $n_0>0$, $0<\gamma<1$ and $\delta >0$ such that for any codebook $\Cc_n, n>n_0$ with $\frac{1}{n} \log|\Cc_n| \geq R$,  we have
	\begin{align}
		P_{e,\rm max}^\metric(\Cc_n,W) \geq  \gamma e^{-\delta} (1 - 2^{- n\bar E(R)}).
	\end{align}
	We set $R = \bar R(W,\metric) + 2\varepsilon$. We know that for any codebook $\Cc_n$ of length $n$ and rate $R$, there exists a constant composition sub-codebook $\Cc'_n\subset\Cc_n$  with length $n$ satisfying, rate $R'>R -  \frac{J-1}{n}\log (n+1)$, and composition $\pv_X$ such that
	\beq \label{gnflsefel}
	P_{e,\rm max}^\metric(\Cc_n,W) \geq P_{e,\rm max}^\metric(\Cc'_n,W). 
	\eeq
	Additionally, from \cite[Lemma 5]{EsiAlbert} for any $\varepsilon>0$ there exists a $\nu>0$ such that there exists a codebook $\tilde \Cc_{\tilde n}$ with the following properties
	\begin{align}
		&\min_{\tilde \pv_X(j)>0} \tilde \pv_X(j) \geq \nu \label{lkjgljhfgpug}\\
		& \tilde{n} \geq n\big(1 - (|\Xc|-1)\nu\big) \label{eq:lemma5eq1}\\
		&P_{e,\rm max}^\metric(\Cc'_n,W) \geq P_{e,\rm max}^\metric(\tilde\Cc_{\tilde n},W) \label{eq:lemma5eq2}\\
		&\frac{1}{\tilde{n}} \log(|\tilde{\Cc}_{\tilde{n}}|) \geq \frac{1}{n}\log(|\Cc'_n|) - \varepsilon + \Oc\Big(\frac{\log n}{n}\Big), \label{eq:lemma5eq3}
	\end{align}	
	where $\tilde \Cc_{\tilde n}$ is of composition $\tilde \pv_X$. Now, let
	\begin{align}
		P_{Y\hat{Y}|X}^\star = \argmin_{\substack{P_{Y\hat{Y}|X} \in \Mmax(\metric, \tilde \pv_{X})\\ P_{Y|X} = W}} I(\tilde \pv_X,P_{\hat{Y}|X})
	\end{align}
	be the best joint conditional distribution for constant composition codes of composition $\tilde \pv_X$.
	Then, by applying Theorem \ref{maintheorem} to $P_{Y\hat{Y}|X}^\star$ we have that there exists a distribution $\bar P_{Y\hat Y|X}$ that fulfills the following conditions 
	\begin{align} \label{gferhfkfmerfe}
		&\pemax(\tilde{\Cc}_{\tilde{n}}, \bar P_{Y|X}) \geq \gamma \petcemax(\tilde{\Cc}_{\tilde{n}},\bar P_{\hat Y|X}). \\
		&|\bar P_{\hat Y|X}\times \tilde \pv_X - P_{\hat Y|X}^\star\times \tilde \pv_X|_{\infty} \leq \frac{1}{\tilde n}  \\
		&|\bar P_{ Y|X}\times \tilde \pv_X - P_{ Y|X}^\star\times \tilde \pv_X|_{\infty} \leq \frac{K}{\tilde n}\\
		&\bar P_{\hat Y|X}\times \tilde \pv_X \in \Pc_{\tilde n}(\Xc\times\Yc)
	\end{align}
	On the other hand, by using Lemma \ref{bcedfkmelfmefe} for $P_{XY}^\star= W\times \pv_{X}$, $\bar P_{XY}= \bar P_{Y|X} \times \pv_{X}$ we have
	\begin{align} \label{bgemfgkfmerf}
		\pemax(\tilde{\Cc}_{\tilde{n}}, P_{Y|X}) \geq e^{-\delta} \pemax(\tilde{\Cc}_{\tilde{n}}, \bar P_{Y|X}),
	\end{align}
	where $\delta = \frac{2K }{\min_{P_{XY}(j,k)>0} P_{XY}(j,k)}.$ 
	To provide an upper bound on $\delta$ note that, we have 
	\begin{align} \label{cekfmefmef}
		P_{XY}^\star(j,k) = W(k|j) \tilde \pv_{X}(j).
	\end{align}
	We observe that since $\tilde \Cc_{\tilde n}$ is a $\nu$-reduction of $\Cc'_n$, from \eqref{lkjgljhfgpug} we have that the right hand side of \eqref{cekfmefmef} is either equal to zero or bigger than or equal to $\nu W(k|j)$. Therefore,
	\begin{align}
		\min_{P_{XY}(j,k)>0} P_{XY}(j,k) \geq \nu\cdot \min_{W(k|j)>0} W(k|j)
	\end{align}
	As a result we have,
	\begin{align}
		0 \leq \delta \leq \frac{2K}{\nu\cdot \min_{W(k|j)>0} W(k|j)},
	\end{align}
	which only depends on the channel, and $\varepsilon$, since $\nu$ depends on $\varepsilon$.
	
	Finally, we apply Theorem \ref{th:3} to codebook $\tilde{\Cc}_{\tilde{n}}$.  Therefore, we have that there exists $n_0$ such that for $n>n_0$ if
	\begin{align}
		\frac{1}{\tilde{n}} \log|\tilde{\Cc}_{\tilde{n}}| &> \max_{P_{X}} \min_{\substack{P_{Y\hat{Y}|X} \in \Mmax(\metric,P_{X})\\ P_{Y|X} = W}} I(P_{X},P_{\hat{Y}|X}) \\
		& \geq \min_{\substack{P_{Y\hat{Y}|X} \in \Mmax(\metric, \tilde \pv_{X})\\ P_{Y|X} = W}} I(\tilde\pv_X,P_{\hat{Y}|X}),\label{dfagahyt}
	\end{align}
	we have that 
	\begin{align} \label{jdimdfedmed}
		\petcemax(\tilde{\Cc}_{\tilde{n}},P_{\hat Y|X}) \geq 1 - 2^{- \tilde n\bar E(R)},
	\end{align}
	where in \eqref{dfagahyt} we have chosen $\tilde \pv_X$ as input distribution instead of the maximizing one.

	Finally, by combining \eqref{gnflsefel}, \eqref{gferhfkfmerfe}, \eqref{bgemfgkfmerf} and \eqref{jdimdfedmed} we get 
	\begin{align}
		\pemax(\Cc_n, P_{Y|X}) &\geq \pemax(\Cc'_n, P_{Y|X}) \\
		&\geq  \pemax(\tilde{\Cc}_{\tilde{n}}, P_{Y|X})  \\
		&\geq e^{-\delta} \pemax(\tilde{\Cc}_{\tilde{n}}, \bar P_{Y|X})\\
		&\geq e^{-\delta} \gamma \petcemax(\tilde{\Cc}_{\tilde{n}},\bar P_{\hat Y|X}) \\
		&\geq  \gamma e^{-\delta} (1 - 2^{- \tilde n\bar E(R)})\\ \label{gerfvacseg}
		&\geq \gamma e^{-\delta} (1 - 2^{- n\,(1 - (|\Xc|-1)\nu)\, \bar E(R)}).
	\end{align}
	
	where \eqref{gerfvacseg} is bounded away from zero as $n$ tends to infinity. 
	
\subsection{Proof of Theorem 2} \label{defoefpesss}

The proof is based on three lemmas. Lemma \ref{bcafkamd}, shows a lower bound to the type-conflict error probability of code $\Cc_n$ over an auxiliary channel. Lemma \ref{abfwfbue}, shows that if the outputs of  $W$ and those of the auxiliary channel and connected by an appropriately constructed graph, then a type-conflict error in the auxiliary channel yields a $\metric$-decoding error in $W$. Lemma \ref{jpiughpkjlASD}, shows that if the joint conditional distribution that defines $W$ and the auxiliary channel is maximal according to \eqref{eq:maximaldef}, then, the error probability of the $\metric$-decoder over channel $W$ is lower-bounded by the type-conflict error probability over the auxiliary channel multiplied by a constant.

\begin{lemma} \label{bcafkamd}
	Assume codebook $\Cc_n$ consists of $M$ codewords of composition $\pv_X$ used over a DMC $P_{\hat Y|X}$. Assume that the conditional type $\pv_{\hat{Y}|X_1}$ is such that $M|\Tc_{\xv_1}( \pv_{\hat Y |X_1}) | \geq 2|\Tc(\pv_{\hat Y})|$. Then, there exists a joint type $\pv_{\hat Y X_1X_2}$ such that $ \pv_{\hat YX_1} = \pv_{\hat YX_2}$ and 
	\begin{align} \label{abdfeidemde}
		&\PP\big[\exists \xv_2 \in \Cc_n \setminus \{\xv_1\} \text{ s.t. } \hat \pv_{\hat \yv \xv_1 \xv_2} =  \pv_{\hat Y X_1X_2}|\xv_1\big] \notag\\
		&~~~~~~~~~~~~~\geq \frac{1}{2(n+1)^{J^2K - 1}} \PP \big[\Tc_{\xv_1}(\pv_{\hat Y |X_1}) |\xv_1\big],
	\end{align} 
	where the probabilities are computed w.r.t. $n$ uses of channel $P_{\hat Y | X}$. 
\end{lemma}
\begin{IEEEproof}
	Before proving this lemma we explain its main application. This lemma implies that at least a polynomial fraction of elements $\hat \yv$ of $\Tc_{\xv_1}(\pv_{\hat \yv | \xv_1})$  cause a type conflict error with some codeword $\xv_2$, when $\xv_1$ is sent and $\hat \yv$ received as the output of the auxiliary channel, for a fixed joint type $\hat \pv_{\hat \yv \xv_1 \xv_2} = \pv_{\hat Y X_1X_2}$.
	
	From \cite[Lemma 4]{gallager_fcc_notes} we have there exist a codeword $\xv_1 \in \Cc_n$ such that
	\begin{align}
		&\PP\big[\exists \xv_2 \in \Cc_n \setminus \{\xv_1\} \text{ s.t. } \hat \pv_{\hat \yv \xv_1} =  \hat \pv_{\hat \yv \xv_2} = \pv_{\hat YX_1}| \xv_1\big]\notag\\
		& ~~~~~~~~~~~~~~~~~~~~~~~~~~~~~~ \geq\frac{1}{2} \PP \big[\Tc_{\xv_1}(\pv_{\hat Y |X_1}) |\xv_1\big] ,
		\label{eq:gallagerlemma}
	\end{align} 
	where the probabilities are computed w.r.t. $n$ uses of channel $P_{\hat Y | X}$. This implies that, assuming $\xv_1\in\Cc_n$ was transmitted, for at least half of the $\hat \yv \in \Tc_{\xv_1}(\pv_{\hat Y |X_1}) $ we can find a codeword $\xv_2\neq \xv_1$ such that $\hat \pv_{\hat \yv | \xv_1} = \hat \pv_{\hat \yv | \xv_2}$. 
	Observe that there are at most $(n+1)^{J^2K-1}$ joint types $\hat \pv_{\hat \yv \xv_1 \xv_2}$. Consider an arbitrary joint type $\tilde \pv_{\hat YX_1X_2}$ and define the subset
	\begin{align}
		&\Ec_{\xv_1}(\tilde\pv_{\hat YX_1X_2},\pv_{\hat YX_1})\notag\\
		&= \big\{ \hat \yv \in \Tc_{\xv_1}( \pv_{\hat Y|X_1}) \,|\, \exists \xv_2 \in \Cc_n\setminus \{\xv_1\},  \notag\\
		&~~~~~~~~~~~\hat \pv_{\hat \yv \xv_1 \xv_2} = \tilde \pv_{\hat Y X_1X_2},  \tilde \pv_{\hat Y X_1}=\tilde \pv_{\hat Y X_2}= \pv_{\hat Y X_1}\big\}.
	\end{align}
	In other words, the set $\Ec_{\xv_1}(\tilde\pv_{\hat YX_1X_2},\pv_{\hat YX_1})$ is the set of outputs $\hat \yv \in \Tc_{\xv_1}( \pv_{\hat Y|X_1})$ such that the joint type of $\yv,\xv_1,\xv_2$ is equal to  $\tilde \pv_{\hat YX_1X_2}$ and the $\hat Y X_1$ and $\hat Y X_2$ marginal types are equal to the given $\pv_{\hat YX_1}$. 
	We now define the joint type $ \pv_{\hat Y X_1 X_2}^\star$ that satisfies the following
	\begin{align}
		\pv_{\hat Y X_1 X_2}^\star = \argmax_{\tilde \pv_{\hat Y X_1 X_2} \in \Pc_n({\Yc\times \Xc^2})} |\Ec_{\xv_1}(\tilde\pv_{\hat YX_1X_2},\pv_{\hat YX_1})|,
		\label{eq:long_type_conflict}
	\end{align}
	i.e., the joint type $\tilde\pv_{\hat Y X_1 X_2}$ that induces the largest subset $\Ec_{\xv_1}(\tilde\pv_{\hat YX_1X_2},\pv_{\hat YX_1})$ for any given $\pv_{\hat YX_1}$. In other words, out of all joint types $\tilde\pv_{\hat Y X_1 X_2}$, $\pv_{\hat Y X_1 X_2}^\star$ is the one that contains the maximum number of outputs $\hat \yv \in \Tc_{\xv_1}( \pv_{\hat Y|X_1})$ that yield a type-conflict error.
	
	Observe that the left hand side of \eqref{eq:gallagerlemma} can be bounded as
	\begin{align}
		\PP\big[\exists &\xv_2 \in \Cc_n \setminus \{\xv_1\} \text{ s.t. } \hat \pv_{\hat \yv \xv_1} =  \hat \pv_{\hat \yv \xv_2} = \pv_{\hat Y X_1 }| \xv_1\big]\notag\\
		&\leq  \sum_{\tilde \pv_{\hat Y X_1 X_2} \in \Pc_n({\Yc\times \Xc^2})} \PP[\Ec_{\xv_1}(\tilde\pv_{\hat YX_1X_2},\pv_{\hat YX_1})| \xv_1] \\
		&\leq (n+1)^{J^2K - 1}\PP[\Ec_{\xv_1}(\pv_{\hat YX_1X_2}^\star,\pv_{\hat YX_1})|\xv_1],
		\label{eq:gallagerlemma2}
	\end{align} 
	where the probability is computed with respect to $n$ uses of channel $P_{\hat Y|X}$, and thus, from \eqref{eq:gallagerlemma}, we get
	\begin{align}
		&\PP\big[\exists \xv_2 \in \Cc_n\setminus \{\xv_1\} \text{ s.t. } \hat \pv_{\hat \yv \xv_1 \xv_2} =  \pv_{\hat Y X_1X_2}^\star| \xv_1\big]\notag\\
		&~~~~~~~~~~~~~~~~~~~ \geq \frac{1}{2(n+1)^{J^2K - 1}} \PP \big[\Tc_{\xv_1}(\pv_{\hat Y |X_1}) |\xv_1\big] 
	\end{align}
	which completes the proof. The joint type $\pv_{\hat Y X_1X_2}^\star$ is the type $\pv_{\hat Y X_1X_2}$ whose existence is stated in the lemma.
\end{IEEEproof}
In the rest of this section whenever $\pv_{\hat Y X_1X_2}^\star$ is used we refer to the type defined in \eqref{eq:long_type_conflict}.
\begin{corollary} \label{xcnanfefefek}
	The above statement implies that
	\begin{align}
		\frac{|\Ec_{\xv_1}(\pv_{\hat YX_1X_2}^\star,\pv_{\hat YX_1})|}{|\Tc_{\xv_1}(\pv_{\hat Y|X_1})|}  \geq \frac{1}{2(n+1)^{J^2K - 1}}. \label{dddvbsefnaodaw22}
	\end{align}
\end{corollary}
\begin{IEEEproof}
	We have that
	\begin{align}
		\frac{|\Ec_{\xv_1}(\pv_{\hat YX_1X_2}^\star,\pv_{\hat YX_1})|}{|\Tc_{\xv_1}(\pv_{\hat Y|X_1})|} \label{hnedfmeofde}
		& = \frac{|\Ec_{\xv_1}(\pv_{\hat YX_1X_2}^\star,\pv_{\hat YX_1})|\cdot\PP[\hat \yv | \xv_1 ] }{|\Tc_{\xv_1}(\pv_{\hat Y|X_1})|\cdot\PP[\hat \yv| \xv_1 ]} \\   \label{chdicjdockepdf}
		& = \frac{\PP\big[\Ec_{\xv_1}(\pv_{\hat YX_1X_2}^\star,\pv_{\hat YX_1})|\xv_1\big]}{\PP \big[\Tc_{\xv_1}(\pv_{\hat Y |X_1}) |\xv_1\big] } \\ \label{dhedaldwldeew}
		& \geq \frac{1}{2(n+1)^{J^2K - 1}},
	\end{align}
	where $\pv_{\hat \yv \xv_1} = \pv_{\hat YX_1}$, \eqref{chdicjdockepdf} follows from the fact that all elements of $\Tc_{\xv_1}(\pv_{\hat Y|X_1})$ are equiprobable when $\xv_1$ is sent and \eqref{dhedaldwldeew} is equivalent to \eqref{abdfeidemde}.
\end{IEEEproof}

Note that in the next lemmas' proof we will employ Corollary \ref{xcnanfefefek} rather than Lemma \ref{bcafkamd}.

Similarly to \cite{EsiAlbert}, we construct a bipartite graph $\Gc_{\xv_1}(\pv_{Y'\hat{Y}|X_1})$ in the following way (see \cite{EsiAlbert} for details). Vertices of this graph are the elements of $\Tc_{\xv_1}(\pv_{Y'|X_1})$ and $\Tc_{\xv_1}(\pv_{\hat Y|X_1})$. Moreover, 
$\yv' \in \Tc_{\xv_1}(\pv_{Y'|X_1})$ and $\hat \yv \in \Tc_{\xv_1}(\pv_{\hat Y|X_1})$ are connected if $\hat \pv_{\yv' \hat \yv \xv_1} = \pv_{Y' \hat Y X_1}$. The graph is regular and we denote the left degree by $d_1$.
Ideally, we need the graph $\Gc_{\xv_1}(\pv_{Y'\hat{Y}|X_1})$ to satisfy the following property: if $\yv'\in \Tc_{\xv_1}(\pv_{Y'|X_1})$ is connected to $\hat \yv \in\Tc_{\xv_1}(\pv_{\hat Y|X_1})$ in this graph then for some $\xv_2 \in \Cc_n / \{\xv_1\}$
\begin{align}
	\metric^n(\xv_2,\yv') \geq \metric^n(\xv_1,\yv').
\end{align}
However, in contrast to \cite{EsiAlbert} this is not always the case here. The next lemma proves a lower bound to the fraction of the edges in $\Gc_{\xv_1}(\pv_{Y'\hat{Y}|X_1})$ that satisfy the aforementioned desired property.

\begin{lemma} \label{abfwfbue}
	Consider a conditional maximal joint type $\pv_{Y' \hat Y |X}\in \Mmaxt(\metric,\pv_X)$, for some composition $\pv_X$, and construct a graph $\Gc_{\xv_1}( \pv_{Y' \hat Y |X_1} )$ between the type classes $\Tc_{\xv_1}(\pv_{\hat Y |X_1})$ and $\Tc_{\xv_1}(\pv_{Y' |X_1})$ as described above. Then, for every $\hat \yv \in \Tc_{\xv_1}(\pv_{\hat Y |X_1})$ such that $\hat \pv_{\hat \yv \xv_1 \xv_2} =  \pv_{\hat Y X_1X_2}^\star$ there are at least $e^{-n\varLambda_n } d_1$ of its neighbours $\yv'\in\Tc_{\xv_1}(\pv_{Y' |X_1})$ such that for some $\xv_2 \in \Cc \backslash \{\xv_1\}$  we have a $\metric$-decoding error when $\xv_1$ is sent, i.e.,
	\begin{align}
		\metricn(\xv_2,\yv') \geq \metricn(\xv_1,\yv'),
	\end{align}
	where  $\varLambda_n =  \Oc \big( \frac{\log n}{n} \big) $.
\end{lemma}

\begin{IEEEproof}
	Consider $\hat \yv \in \Tc_{\xv_1}(\pv_{\hat Y |X_1})$. By construction, all $\yv'$ that are connected to $\hat \yv$ in graph $\Gc_{\xv_1}( \pv_{Y' \hat Y |X_1} )$ satisfy 
	\begin{align}
		\hat \pv_{\yv' \hat \yv | \xv_1} = \pv_{Y' \hat Y |X}.
	\end{align}	
	As a result, by using  Lemma \ref{vrogjmfiarnqawen} with $T = Y', S = X_2$, $Z = (\hat Y,X_1)$ and $f(T,S) = \metric(X_2, Y')$ we have
	\begin{align} \label{caokeitjiezfme}
		\EE_U[\metric^n(\xv_2,\yv')] &= n\EE_{\pv_{Y'|\hat Y X_1} \times \pv_{\hat Y X_1 X_2}^\star }[\metric(X_2,Y')] \\ \label{bedijmemdw}
		&\geq n\EE[q(X_1,Y')] \\ \label{nrfofepdanfgjfe}
		&= \metric^n(\xv_1,\yv'),
	\end{align}
	where $U$ is a equiprobable random variable over all sequences $\yv'\in \Tc_{\xv_1}(\pv_{Y' |X_1})$ that satisfy $\hat \pv_{\yv' \hat \yv \xv_1} = \pv_{Y' \hat Y X_1}$ and $\hat \pv_{\hat \yv \xv_1\xv_2} = \pv_{\hat Y X_1 X_2}^\star$, \eqref{bedijmemdw} follows from $\pv_{Y'\hat Y |X_1}$ being maximal and \eqref{nrfofepdanfgjfe} is derived from the additivity of the metric $\metric$.
	
	The above equation has an important implication: the expected metric computed on all  sequences $\yv'$ that satisfy $\hat \pv_{\yv' \hat \yv \xv_1} = \pv_{Y' \hat Y X_1}$ is larger than or equal to $\metric^n(\xv_1,\yv')$. In the rest of this section, we derive a subset of all such $\yv'$ that satisfy the mismatch pairwise error condition $\metric^n(\xv_2,\yv') \geq \metric^n(\xv_1,\yv')$.
	
	The main difficulty in deducing finding such a set directly from \eqref{nrfofepdanfgjfe} is that $\pv_{Y'|\hat YX_1} \times \pv_{\hat Y X_1 X_2}^\star $ might not be a type. As a result, there might not be any $\yv'$ with the type $\pv_{Y'|\hat Y X_1} \times \pv_{\hat Y X_1 X_2}^\star $ which satisfies the desired inequality \eqref{nrfofepdanfgjfe}. Therefore, we attempt to express this distribution $\pv_{Y'|\hat Y X_1} \times \pv_{\hat Y X_1 X_2}^\star $ as a linear combination of types that are in a neighborhood of $\pv_{Y'|\hat Y X_1} \times \pv_{\hat Y X_1 X_2}^\star $, and then prove the desired property for one such type.

	By using Lemma \ref{bogjamirenruanr} with $Z= Y',S=(\hat Y,X_1),U  = X_2$, respectively, we can express the distribution $\pv_{Y'|\hat Y X_1} \times \pv_{\hat Y X_1 X_2}^\star$ as a convex combination of joint types $\tilde \pv_{Y'\hat YX_1X_2}$ with marginals $\pv_{Y'\hat Y X_1}$ and $\pv_{\hat Y X_1 X_2}^\star$ for which $|\tilde \pv_{Y'\hat Y X_1X_2} - \pv_{Y'|\hat Y X_1} \times \pv_{\hat Y X_1 X_2}^\star|_{\infty} \leq \frac{1}{n}$. More precisely, we have
	\begin{align}\label{vdfmizmfeaiof}
		&\pv_{Y'|\hat Y X_1} \times \pv_{\hat Y X_1 X_2}^\star \notag\\
		&= \sum_{ \substack{            \tilde \pv_{Y'\hat Y X_1X_2} \\ |\tilde \pv_{Y'\hat Y X_1X_2} - \pv_{Y'|\hat YX_1} \times \pv_{\hat Y X_1 X_2}^\star|_{\infty} \leq \frac{1}{n} \\ \tilde \pv_{Y'\hat Y X_1} = \pv_{Y'\hat Y X_1} , \tilde \pv_{\hat Y X_1 X_2} = \pv_{\hat Y X_1 X_2}^\star    }  }  \hspace{-5mm}\alpha(\tilde \pv_{Y'\hat YX_1X_2}) \tilde \pv_{Y'\hat Y X_1X_2},
	\end{align}
	where
	\begin{align}
		\sum_{ \substack{            \tilde \pv_{Y'\hat YX_1X_2} \\ |\tilde \pv_{Y'\hat Y X_1X_2} - \pv_{Y'|\hat YX_1} \times \pv_{\hat Y X_1 X_2}^\star|_{\infty} \leq \frac{1}{n} \\ \tilde \pv_{Y'\hat Y X_1} = \pv_{Y'\hat Y X_1} , \tilde \pv_{\hat Y X_1 X_2} = \pv_{\hat Y X_1 X_2}^\star    }  }  \hspace{-5mm}\alpha(\tilde \pv_{Y'\hat YX_1X_2}) = 1 
	\end{align}
	and $\alpha(\tilde \pv_{Y'\hat YX_1X_2}) \geq 0$.

	Therefore, from \eqref{caokeitjiezfme} and \eqref{vdfmizmfeaiof} we have
	\begin{align}
		&\EE_U[\metric(\xv_2,\yv') ] \notag\\[7pt]
		&= n\EE_{\pv_{Y'|\hat YX_1} \times \pv_{\hat Y X_1 X_2}^\star }[\metric(X_2,Y')] \\
		&= n \sum_{ \substack{ \tilde \pv_{Y'\hat YX_1X_2} \\ |\tilde \pv_{Y'\hat YX_1X_2} - \pv_{Y'|\hat YX_1} \times \pv_{\hat Y X_1 X_2}^\star|_{\infty} \leq \frac{1}{n} \\ \tilde \pv_{Y'\hat YX_1} = \pv_{Y'\hat Y X_1} , \tilde \pv_{\hat Y X_1 X_2} = \pv_{\hat Y X_1 X_2}^\star    }  }  \hspace{-5mm}\alpha(\tilde \pv_{Y'\hat YX_1X_2})\notag\\[10pt]
		&~~~~~~~~~~~~~~~~~~~~~~~~~~~~~~~~~~~~\times  \EE_{\tilde \pv_{Y'\hat Y X_1X_2}} [\metric(\xv_2,\yv')].
	\end{align}
	Moreover, from \eqref{nrfofepdanfgjfe} we know $\EE_U[\metric^n(\xv_2,\yv')] \geq \metric^n(\xv_1,\yv') $,  and therefore, there exists a joint type $\tilde \pv_{Y'\hat Y X_1X_2}$ such that 
	
	\begin{align}
		n \EE_{\tilde \pv_{Y'\hat Y X_1X_2} } [\metric(X_2,Y') ] &= \metric^n(\xv_2,\yv')\\
		&\geq \metric^n(\xv_1,\yv').\label{dafhhj}
	\end{align}
	As a result, by using $|\tilde \pv_{Y'\hat YX_1X_2} - \pv_{Y'|\hat YX_1} \times \pv_{\hat Y X_1 X_2}^\star|_{\infty} \leq \frac{1}{n}$ and Lemma \ref{odeormewr} we obtain a lower bound on the number of $\yv'$ with   the above type $\hat \pv_{\yv' \hat \yv \xv_1 \xv_2} = \tilde \pv_{Y'\hat Y X_1X_2}$
	 \begin{align}\label{pfrfpedkexdxxx}
	& |\{ \yv' \in \Tc_{\xv_1 \hat{\yv}}( \pv_{Y'|\hat{Y}X_1})| \exists \xv_2 \in \Cc_{n} \backslash \{\xv_1\},\notag\\
	 &~~~~~~~~~~~~~~~ \hat \pv_{\yv' \hat \yv \xv_1 \xv_2 } = \tilde \pv_{Y'\hat Y X_1X_2} \}| = e^{n H(Y'|\hat Y, X_1)  - n\omega_n}.
	 \end{align}
	where the entropy is computed using probability distribution $\pv_{Y'|\hat Y X_1} \times \pv_{\hat Y X_1 X_2}^\star$.
	
	On the other hand, since $d_1$ is defined as degree of every $\hat \yv \in \Tc_{\xv_1}(\pv_{\hat Y|X_1})$ in graph $\Gc_{\xv_1}(\pv_{Y'\hat{Y}|X_1})$ we have that $d_1=e^{n H(Y'|\hat Y, X_1) - n\kappa_n } $ where $\kappa_n = \Oc \big( \frac{\log n}{n} \big)$. This follows from the type counting lemma from Gallager's notes \cite{gallager_fcc_notes} and it can be derived by noting that degree $d_1$ is equal to the number of sequences $\yv'$ such that $\hat \pv_{\hat \yv \yv' \xv_1} = \pv_{\hat Y Y' X_1}$ when $\xv_1, \hat \yv$ are fixed, more precisely $d_1 = |\Tc_{\xv_1 \hat \yv}(P_{ Y' |\hat Y X_1})|$.
	
	As a result, by combining \eqref{pfrfpedkexdxxx} and the fact that $d_1 = e^{n H(Y'|\hat Y, X_1) - n\kappa_n}$ we have 
	\begin{align}
		&|\{ \yv' \in \Tc_{\xv_1\hat \yv}(\pv_{Y'\hat{Y}|X_1}) | \exists \xv_2 \in \Cc_{n} \backslash \{\xv_1\},\notag\\
		&~~~~~~~~~~~~~~~~~ \hat \pv_{\yv' \hat \yv \xv_1 \xv_2} = \pv_{Y'\hat Y X_1X_2}^\star \}| = e^{- n\omega_n + n\kappa_n} \cdot d_1
		\label{fsgw6}
	\end{align}
	Also, from  \eqref{dafhhj}, for every $\yv'$ in the above set \eqref{fsgw6}, we have
	\begin{align}
		\metric^n(\xv_2,\yv') \geq \metric^n(\xv_1,\yv').
	\end{align}
	By setting $\varLambda_n = \omega_n - \kappa_n$ we get the desired result.
\end{IEEEproof}
Now we construct a new graph $\tilde \Gc_{\xv_1}(\pv_{Y'\hat{Y}|X_1})$ using Lemma \ref{abfwfbue}. We construct this graph by starting from $\Gc_{\xv_1}(\pv_{Y'\hat{Y}|X_1})$ and for each $\hat \yv \in \Tc_{\xv_1}(\pv_{\hat Y |X_1})$ only keeping the edges that are connected to $\yv'$ that for some $\xv_2 \in \Cc \backslash \{\xv_1\}$  we have a $\metric$-decoding error, more precisely
\begin{align} \label{cbodwakdwkodfr}
	\metricn(\xv_2,\yv') \geq \metricn(\xv_1,\yv').
\end{align}
As described in \cite{EsiAlbert}, the graph $\Gc_{\xv_1}(\pv_{Y'\hat{Y}|X_1})$ is regular: for every $\yv' \in \Tc_{\xv_1}(\pv_{Y'|X_1})$ the number of $\hat \yv \in \Tc_{\xv_1}(\pv_{\hat Y|X_1})$ such that $\hat \pv_{\yv' \hat \yv \xv_1} = \pv_{Y' \hat Y X_1}$ is the same; similarly, for every $\hat\yv \in \Tc_{\xv_1}(\pv_{\hat Y|X_1})$ the number of $\yv' \in \Tc_{\xv_1}(\pv_{Y'|X_1})$ such that $\hat \pv_{\yv' \hat \yv \xv_1} = \pv_{Y' \hat Y X_1}$ is the same.

The graph $\tilde \Gc_{\xv_1}(\pv_{Y'\hat{Y}|X_1})$ is no longer regular. The previous lemma shows that the degree of any vertex in $\tilde \Gc_{\xv_1}(\pv_{Y'\hat{Y}|X_1})$ is at least $e^{-n\varLambda_n } d_1$ and $\varLambda_n =  \Oc \big( \frac{\log n}{n} \big) $. Now we can use this fact to prove the next lemma which relates the $\metric$-decoding error probability in channel $P_{Y'|X}$ with the type-conflict error probability in channel $P_{\hat Y|X}$.
\begin{lemma}\label{jpiughpkjlASD}
	Let $\pv_{Y'\hat{Y}|X_1} \in \hat{\Mc}_{\max}(\metric, \pv_{X_1})$ be a maximal joint conditional type and $\xv_1 \in \Tc(\pv_{X_1})$ be the transmitted codeword. Then
	\begin{align}
		\pemax(\Cc_n,W) \geq e^{-n\sigma_n} \PP [\Tc_{\xv_1}(\pv_{ Y' |X_1}) |\xv_1],
	\end{align}
	where $\sigma_n =  \Oc \big( \frac{\log n}{n} \big) $ and both probabilities are computed with respect to $n$ uses of channel $W$.
\end{lemma}
\begin{IEEEproof}
	Consider the bipartite graph $\tilde \Gc_{\xv_1}(\pv_{Y'\hat{Y}|X_1})$ obtained by connecting elements $\Tc_{\xv_1}(\pv_{Y'|X_1})$ with $\Tc_{\xv_1}(\pv_{\hat Y|X_1})$ as described above. For any $\Bc \subset \Tc_{\xv_1}(\pv_{\hat Y|X_1})$ we define $\Psi(\Bc)$ as
	\begin{align}
		\Psi(\Bc)&= \big\{\yv' \in \Tc_{\xv_1}(\pv_{Y'|X_1}) \ |\ \yv' \text{ is connected}\notag\\
		&~~~~~~~~~~~\text{ to some } \hat \yv \in \Bc \text{ in graph } \tilde \Gc_{\xv}(\pv_{Y'\hat{Y}|X_1}) \big\}
	\end{align} 
	We apply Lemma \ref{graphlemma} to graph $\tilde \Gc_{\xv_1}(\pv_{Y'\hat{Y}|X_1})$ and we obtain that for any $ \Bc \subset \Tc_{\xv_1}(\pv_{\hat Y|X_1})$
	\begin{align} \label{vsojgoaef}
		\frac{|\Psi(\Bc)|}{|\Tc_{\xv_1}(\pv_{ Y'|X_1})|} \geq e^{-n\varLambda_n } \frac{|\Bc|}{|\Tc_{\xv_1}(\pv_{\hat Y|X_1})|}.
	\end{align}
	Now, let $\Bc$ be the set of all $\hat \yv\in\Tc_{\xv_1}(\pv_{\hat Y|X_1})$ such that there exist a type-conflict error with another codeword $\xv_2$ such that $\hat \pv_{\hat \yv \xv_1 \xv_2} = \pv_{\hat Y X_1X_2}^\star$ from Lemma \ref{bcafkamd} Eq. \eqref{eq:long_type_conflict}, i.e., 
	\beq
	\Bc = \Ec_{\xv_1}(\pv_{\hat YX_1X_2}^\star,\pv_{\hat YX_1}).
	\eeq
	Therefore, from Lemma \ref{abfwfbue} we have for any $\yv' \in \Psi(\Bc)$ there exists a codeword $\xv_2 \neq \xv_1$ such that
	\begin{align} \label{dvawhfja}
		\metricn(\xv_2, \yv') \geq \metricn(\xv_1,\yv').
	\end{align}	
	We bound the probability of error as follows 
	\begin{align}
		&\pemax(\Cc_n,W)\notag\\
		&= \PP [\exists \xv_2 \in \Cc_n \backslash \{\xv_1\}, \metricn(\xv_2,\yv') \geq \metricn(\xv_1,\yv')| \xv_1] \\
		&\geq  \PP [\exists \xv_2 \in \Cc_n \backslash \{ \xv_1 \}, \metricn(\xv_2,\yv') \geq \metricn(\xv_1,\yv'),\notag\\
		&~~~~~~~~~~~~~~~~~~~~~~~~~~~~~~~~~~~~~\yv' \in \Tc_{\xv_1}(\pv_{Y'|X_1}) |\xv_1 ] \\
		&=  \PP[\Tc_{\xv_1}(\pv_{Y'|X_1})|\xv_1] \cdot \PP [\exists \xv_2 \in \Cc_n \backslash \{ \xv_1 \}, \notag\\
		&~~~~~~~~\metricn(\xv_2,\yv') \geq \metricn(\xv_1,\yv')|\yv' \in \Tc_{\xv_1}(\pv_{Y'|X_1}) , \xv_1 ] \\
		& = \PP[\Tc_{\xv_1}(\pv_{Y'|X_1})|\xv_1]\cdot \\
		&\frac{\big|\mspace{-2mu}\big\{\yv' \mspace{-5mu} \in \mspace{-5mu} \Tc_{\xv_1}(\pv_{Y'|X_1}) | \exists \xv_2 \mspace{-5mu} \in \mspace{-5mu} \Cc_n \backslash \{ \xv_1 \}, \metricn( \xv_2,\yv')\mspace{-5mu} \geq \mspace{-5mu} \metricn(\xv_1,\yv')\big\}\mspace{-2mu}\big|}{|\Tc_{\xv_1}(\pv_{Y'|X_1})|} \\ \label{psi}
		&\geq \PP[\Tc_{\xv_1}(\pv_{Y'|X_1})|\xv_1]\cdot \frac{\big|\Psi\big(\Ec_{\xv_1}(\pv_{\hat YX_1X_2},\pv_{\hat YX_1})\big)\big|}{|\Tc_{\xv_1}(\pv_{ Y'|X_1})|}\\ \label{amfnmwiti}
		&\geq \PP[\Tc_{\xv_1}(\pv_{Y'|X_1})|\xv_1]\cdot e^{-n\varLambda_n } \cdot \frac{|\Ec_{\xv_1}(\pv_{\hat YX_1X_2},\pv_{\hat YX_1})|}{|\Tc_{\xv_1}(\pv_{\hat Y|X_1})|} \\
		&\geq \PP[\Tc_{\xv_1}(\pv_{Y'|X_1})| \xv_1]\cdot e^{-n\varLambda_n } \frac{1}{2(n+1)^{J^2K - 1}}, \label{bysaefbnwief}
	\end{align}
	where all of probabilities are computed with respect to $n$ uses of channel $W$,  \eqref{psi} follows from all elements of $\Psi(\Bc)$ satisfying \eqref{dvawhfja}, \eqref{amfnmwiti} follows from \eqref{vsojgoaef} and \eqref{bysaefbnwief} follows from \eqref{dddvbsefnaodaw22}. By setting $\delta_n = \varLambda_n + (J^2K - 1)\frac{\log (n+1)}{n}$ we get the desired result.
\end{IEEEproof}

Using a standard property of conditional types we have that
\beq
\PP[\Tc_{\xv_1}(\pv_{Y'|X_1})| \xv_1] \geq e^{-n \big(D(P_{Y'|X_1} \| P_{Y|X_1} | \pv_{X_1} ) +\delta_n\big) }
\eeq
with $\delta_n = \Oc \big( \frac{\log n}{n} \big)$.  From standard arguments of the method of types we obtain \eqref{eq:spb}, where we have set $\pv_X=\pv_{X_1}$.

Again using standard arguments (see e.g. \cite[Th. 2]{gallager_fcc_notes}) the result of Theorem \ref{dafbeghr} is applicable to any code, and not only constant composition codes. This is due to the fact that every codebook $\Cc_n$ of rate $R$ has a constant composition sub-codebook $\Cc_n'\subseteq\Cc_n$ with rate $R'>R -  \frac{J-1}{n}\log (n+1)$ with 
\beq
\pemax(\Cc_n,W) \geq \pemax( \Cc_n',W).
\label{eq:pemaxboundcc}
\eeq 
%
%
Additionally, a similar analysis would give an identical upper bound to the error exponent using the maximal sets $\Mmaxt(\metric)$ from \cite{EsiAlbert}. 

As is well known, the exponent from Theorem \ref{dafbeghr} is decreasing in $R$ and $\Espq(\pv_X,R) = 0$ by choosing $Y' = Y$ in \eqref{vnsefiaefieafn} at a rate equal to
\begin{align}
	\bar R_\metric(W,\pv_X) \triangleq \min_{\substack{P_{Y\hat{Y}|X} \in \Mmax(\metric, \pv_{X})\\ P_{Y|X} = W}} I(\pv_X,P_{\hat{Y}|X})
\end{align}
We have shown that for rates $R<\bar R_\metric(W,\pv_X)$, the error probability decays at most exponentially. The proof of Theorem \ref{th:main2} in Section \ref{fepfepdkwea} shows that for rates $R> \max_{P_X }\bar R_\metric(W,P_X)$ the error probability cannot decay sub-exponentially and is bounded away from zero as $n$ tends to infinity. In the next subsection, we extend our error exponent analysis to type-dependent metrics.

\subsection{Type-Dependent Metrics}
\label{sec:tdm}
In this part we show the previous analysis holds for an important family of type-dependent metrics as well. Namely, we show the analysis holds for type-dependant metric $\metric(P_{XY})$ where $\metric$ is convex in $P_{Y|X}$ when $P_X$ is fixed. This is an important family since important metrics such as maximum mutual information (MMI) metric defined as $\metric(P_{XY}) = I(P_{XY})$ have this property. With a slight abuse of notation we use $\metric(\pv_{XY})$ to denote a type-dependent metric $\metric$ computed for type $\pv_{XY}$.  Recall definition of $\Mmax^{\rm td}(\metric,P_{X})$ 
	\begin{align}
		&\Mmax^{\rm td}(\metric,P_{X_1}) \notag\\
		&~~~~\eqdef \Bigg \{ P_{Y\hat Y|X_1} \bigg| \min_{\substack{P_{X_2|X_1 \hat Y}: \\ X_2- \hat Y X_1 - Y \\ P_{\hat Y X_1} = P_{\hat Y X_2} }}  \metric(P_{X_2Y}) \geq \metric(P_{X_1Y}) \Bigg \}.
	\end{align}

For this family of metrics, we have exactly the same statement as that of Theorem \ref{dafbeghr}, but replacing $\Mmax(\metric,P_{X_1})$ by $\Mmax^{\rm td}(\metric,P_{X_1})$.

Here we only discuss the parts of the proof that are different from that of Theorem \ref{dafbeghr}.
To begin with, let $\pv_{Y' \hat Y |X}\in \Mmaxt(\metric,\pv_X)$. Lemma \ref{bcafkamd} remains valid since the result and its proof do not depend on the decoding metric nor its form. We now adapt Lemma \ref{abfwfbue} to type-dependent metrics. Assume, we have $\pv_{\hat Y X_1X_2}^\star$ as explained in the proof of the Lemma \ref{bcafkamd}. Moreover, the graph $\Gc_{\xv_1}( \pv_{Y' \hat Y |X_1} )$ is constructed similarly. We now want to construct a graph $\tilde \Gc_{\xv_1}( \pv_{Y' \hat Y |X_1} )$ analogously to the proof of Lemma \ref{abfwfbue}. To this end, by using Lemma \ref{bogjamirenruanr} with $Z,S,U = Y', (\hat Y,X_1), X_2$, respectively, we can express $\pv_{Y'|\hat Y,X_1} \times \pv_{\hat Y X_1 X_2}^\star$ as a convex combination of types that have marginals equal to $\pv_{Y'\hat Y,X_1}, \pv_{\hat Y X_1 X_2}^\star$ and satisfying $|\tilde \pv_{Y'\hat Y,X_1X_2} - \pv_{Y'|\hat Y,X_1} \times \pv_{\hat Y X_1 X_2}^\star|_{\infty} \leq \frac{1}{n}$. More precisely, we have
\begin{align}\label{mcifnrbfgrbdd}
	&\pv_{Y'|\hat Y,X_1} \times \pv_{\hat Y X_1 X_2}^\star \notag\\
	&= \sum_{ \substack{            \tilde \pv_{Y'\hat Y,X_1X_2} \\ |\tilde \pv_{Y'\hat Y,X_1X_2} - \pv_{Y'|\hat Y,X_1} \times \pv_{\hat Y X_1 X_2}^\star|_{\infty} \leq \frac{1}{n} \\ \tilde \pv_{Y'\hat Y,X_1} = \pv_{Y'\hat Y,X_1} , \tilde \pv_{\hat Y X_1 X_2} = \pv_{\hat Y X_1 X_2}^\star    }  } \hspace{-5mm} \alpha(\tilde \pv_{Y'\hat YX_1X_2}) \tilde \pv_{Y'\hat Y,X_1X_2},
\end{align}
where 
\begin{align}
	\sum_{ \substack{            \tilde \pv_{Y'\hat YX_1X_2} \\ |\tilde \pv_{Y'\hat Y,X_1X_2} - \pv_{Y'|\hat YX_1} \times \pv_{\hat Y X_1 X_2}^\star|_{\infty} \leq \frac{1}{n} \\ \tilde \pv_{Y'\hat Y,X_1} = \pv_{Y'\hat Y,X_1} , \tilde \pv_{\hat Y X_1 X_2} = \pv_{\hat Y X_1 X_2}^\star    }  }  \hspace{-5mm}\alpha(\tilde \pv_{Y'\hat YX_1X_2}) = 1 
\end{align}
and $\alpha(\tilde \pv_{Y'\hat YX_1X_2}) \geq 0$. We will now show that there exists a type $\pv_{Y'\hat Y X_1X_2}^\star$ such that $|\tilde \pv_{Y'\hat Y,X_1X_2} - \pv_{Y'|\hat Y,X_1} \times \pv_{\hat Y X_1 X_2}^\star|_{\infty} \leq \frac{1}{n}$ and the mismatched decoder makes an error, i.e.,
\begin{align}
	\metric(\pv_{YX_2}) \geq \metric(\pv_{YX_1}).
\end{align}
This can be seen by the fact that $\metric(P_{XY})$ is convex in $P_{Y|X}$ and using \eqref{mcifnrbfgrbdd}. More precisely, if we define $f(P_{Y\hat YX_1X_2}) = \metric(P_{X_2Y})$, then, $f(P_{Y\hat YX_1X_2})$ is convex in $P_{Y\hat YX_1X_2}$ when $P_{X_2}$ is fixed. As a result, we have
\begin{align}
	&\metric(P_{Y'X_2}) \notag\\
	&= f(\pv_{Y'|\hat Y,X_1} \times \pv_{\hat Y X_1 X_2}^\star) \\ \label{cbdeemfekfrgrg}
	&= f \Bigg (\hspace{-3mm}\sum_{ \substack{            \tilde \pv_{Y'\hat Y,X_1X_2} \\ |\tilde \pv_{Y'\hat Y,X_1X_2} - \pv_{Y'|\hat Y,X_1} \times \pv_{\hat Y X_1 X_2}^\star|_{\infty} \leq \frac{1}{n} \\ \tilde \pv_{Y'\hat Y,X_1} = \pv_{Y'\hat YX_1} , \tilde \pv_{\hat Y X_1 X_2} = \pv_{\hat Y X_1 X_2}^\star    }  }  \hspace{-7mm}\alpha(\tilde \pv_{Y'\hat YX_1X_2}) \tilde \pv_{Y'\hat Y,X_1X_2} \Bigg) \\ \label{cdcndjcnedeodkeo}
	&\leq \sum_{ \substack{            \tilde \pv_{Y'\hat Y,X_1X_2} \\ |\tilde \pv_{Y'\hat Y,X_1X_2} - \pv_{Y'|\hat Y,X_1} \times \pv_{\hat Y X_1 X_2}^\star|_{\infty} \leq \frac{1}{n} \\ \tilde \pv_{Y'\hat Y,X_1} = \pv_{Y'\hat Y,X_1} , \tilde \pv_{\hat Y X_1 X_2} = \pv_{\hat Y X_1 X_2}^\star    }  }  \hspace{-7mm}\alpha(\tilde \pv_{Y'\hat YX_1X_2}) f(\tilde \pv_{Y'\hat Y,X_1X_2}) \\
	& = \sum_{ \substack{            \tilde \pv_{Y'\hat Y,X_1X_2} \\ |\tilde \pv_{Y'\hat Y,X_1X_2} - \pv_{Y'|\hat Y,X_1} \times \pv_{\hat Y X_1 X_2}^\star|_{\infty} \leq \frac{1}{n} \\ \tilde \pv_{Y'\hat Y,X_1} = \pv_{Y'\hat YX_1} , \tilde \pv_{\hat Y X_1 X_2} = \pv_{\hat Y X_1 X_2}^\star    }  }  \hspace{-7mm}\alpha(\tilde \pv_{Y'\hat YX_1X_2}) \metric(\tilde \pv_{Y'X_2}),
\end{align}
where \eqref{cbdeemfekfrgrg} follows by substituting \eqref{mcifnrbfgrbdd}, and \eqref{cdcndjcnedeodkeo} follows by the convexity of $f$. Throughout this derivation $\pv_{X_2}$ is fixed due to the codebook being of constant composition. Therefore, we obtain the desired result. 
We now use this type $\pv_{Y'\hat Y X_1 X_2}^\star$ to construct the graph $\tilde \Gc_{\xv_1}( \pv_{Y' \hat Y |X_1} )$. The proof proceeds as that of Theorem \ref{dafbeghr} with the new graph $\tilde \Gc_{\xv_1}( \pv_{Y' \hat Y |X_1} )$.

\subsubsection*{Proof of Theorem \ref{th:3}}

The proof is almost identical to the proof of Theorem \ref{dafbeghr}; we point out the steps that are different. The main difference with the proof of Theorem \ref{dafbeghr} is the order of the choice of the type $\pv_{\hat YX_1X_2}^\star$ and choice of the maximal joint conditional type as mentioned above. Suppose that we fix $\pv_{Y'|X}$ and $\pv_{\hat Y|X}\in \hat\Vc_{\rm max}^{\rm td}(\metric,P_{X_1},\pv_{Y'|X})$ and consider the type classes $\Tc_{\xv_1}(\pv_{\hat Y |X_1}), \Tc_{\xv_1}(\pv_{Y' |X_1})$. Note that we cannot yet construct a graph between these two type classes, because we have not specified a joint conditional type. Yet,  Lemma \ref{bcafkamd} still holds, since the maximality condition of the underlying joint conditional type is not used. Consider the joint type $\pv_{\hat YX_1X_2}^\star$ from Lemma \ref{bcafkamd}.
We can now use the definition of $\hat\Vc_{\rm max}^{\rm td}(\metric,P_{X},\pv_{Y'|X} )$ for which
	\begin{align} \label{cndimeidef}
		\max_{\substack{P_{Y\hat Y |X_1}: \\ P_{Y X_1} = \pv_{Y'X_1}\\ P_{\hat Y X_1} = \pv_{\hat YX_1}  }} \min_{\substack{P_{X_2|X_1 \hat Y}: \\ X_2- \hat Y X_1 - Y \\ P_{\hat Y X_1} = P_{\hat Y X_2} }}  \metric(P_{X_2Y}) \geq \metric(P_{X_1Y}).
	\end{align}
	In other words, there exists a joint conditional distribution $P_{\hat Y Y' |X_1}$ which is the maximizer in \eqref{cndimeidef} such that the marginals satisfy
	\begin{align}
		&P_{ Y' X_1} = \pv_{Y' X_1} \\
		&P_{\hat Y X_1} = \pv_{\hat Y X_1},
	\end{align}
	and additionally,
	\begin{align}
	 \metric(P_{X_2Y}) \geq \metric(P_{X_1Y}),
	\end{align}
	where $X_1,X_2, Y', \hat Y \sim P_{Y' |\hat Y  X_1}\pv_{\hat Y X_1 X_2}^\star$. The proof proceeds as in Section \ref{sec:tdm}.

\appendices
\section{Proof of Theorem \ref{maintheorem}} \label{auxiliary_theorems}
In this section, we prove Theorem \ref{maintheorem}. The proof uses several results stated and proved in Appendices \ref{useful_lemmas} and \ref{xvajfefodefmer}.

	We first show existence of a joint conditional distribution $\bar P_{Y\hat Y |X} \in \Mmax(q, \pv_X)$ with properties \eqref{eq:pbar1}, \eqref{eq:pbar2} and \eqref{eq:pbar3}.
	To  this end let the joint distribution be $ P_{Y\hat Y X} = P_{Y\hat Y |X} \times \pv_X$. Then, we can use Lemma \ref{oefkznfdwarnawe} to express $P_{X\hat Y}$ as follows
	\begin{align}
		P_{X\hat Y } = \sum_{\substack{P'_{\hat Y X}  \in \Pc_n(\Xc\Yc): \\  |P'_{\hat Y X} - P_{\hat Y X}|_\infty \leq \frac{1}{n} }} \alpha(P'_{X\hat Y})  P'_{X\hat Y },
		\label{eq:lkghadfg}
	\end{align}
	where the coefficients $\alpha(\cdot)$ are non-negative and are such that $\sum_{\substack{P'_{\hat Y X} \in \Pc_n(\Xc\Yc): \\  |P'_{\hat Y X} - P_{\hat Y X}|_\infty \leq \frac{1}{n} }} \alpha(P'_{X\hat Y})  = 1$.
	As a result, by multiplying both sides of \eqref{eq:lkghadfg} by $P_{Y|\hat Y X_1}$ we have that
	\begin{align}
		P_{Y\hat Y X} = \sum_{\substack{P'_{\hat Y X} \in \Pc_n(\Xc\Yc) : \\  |P'_{\hat Y X} - P_{\hat Y X}|_\infty \leq \frac{1}{n} }} \alpha(P'_{\hat Y X}) P_{Y| \hat Y X}  P'_{\hat Y X}.
		\label{sgadfga}
	\end{align}
	Define the joint conditional distributions 
	\beq
	P'_{Y\hat Y X} = P_{Y| \hat Y X}  P'_{\hat Y X}
	\label{eq:sghsgh}
	\eeq
	in the sum \eqref{sgadfga}. 
	
	The theorem statement assumes that $P_{Y\hat Y |X}$ is maximal. 
	We now claim that at least one of the joint conditional distributions $P'_{Y\hat Y X} = P_{Y| \hat Y X}  P'_{\hat Y X}$ in the sum \eqref{sgadfga} is maximal. To see this, assume by contradiction none of the joint conditional distributions $P'_{Y\hat Y X} = P_{Y| \hat Y X}  P'_{\hat Y X}$ in the sum \eqref{sgadfga} are maximal. This  this implies that for each distribution $P'_{Y\hat Y X_1}$ there exists a 
	distribution $P_{X_2|X_1\hat Y}^\star$ such that the optimization problem in the definition of the maximal set gives
	\begin{align}
		P_{X_2|X_1\hat Y}^\star	=\argmin_{\substack{P_{X_2|X_1\hat Y}: \\ X_2-X_1\hat Y-Y \\P'_{\hat{Y}X_2} = P'_{\hat{Y}X_1}}} \EE [\metric(X_2,Y)] < \EE [\metric(X_1,Y)],
		\label{eq:maximaldefopt}
	\end{align}
	where the expectations in \eqref{eq:maximaldefopt} are computed over joint distributions $P'_{Y\hat Y X_1X_2} = P_{X_2|X_1\hat Y}^\star P'_{Y\hat Y X_1}$.
	
	Define
	\begin{align} \label{cnfralfefmefr}
		P_{Y\hat Y X_1 X_2} = \sum_{\substack{P'_{\hat Y X} \in \Pc_n(\Xc\Yc): \\  |P'_{\hat Y X} - P_{\hat Y X}|_\infty \leq \frac{1}{n} }} \alpha(P'_{\hat Y X}) P'_{Y\hat YX_1X_2}
	\end{align}
	with the same coefficients as in \eqref{eq:lkghadfg}.
	We have that	
	\begin{align} \label{cnefelgfmert}
		P_{\hat Y X_1} &= P_{\hat Y X_2}\\ \label{fhefnerfmerf}
		P_{Y\hat Y X_1 X_2} &= P_{Y|\hat YX_1} P_{\hat Y X_1X_2},
	\end{align}
	where \eqref{cnefelgfmert} follows from the fact that all $P'_{\hat YX_1X_2}$ in the sum of equation \eqref{cnfralfefmefr} are such that $P'_{\hat Y X_1} = P'_{\hat Y X_2}$ by construction and \eqref{fhefnerfmerf} follows from the definition  of $P'_{Y\hat Y X}$ in \eqref{eq:sghsgh}.
	
	We write the expectation condition in \eqref{eq:maximaldefopt} as
	\begin{align}\label{fjegfmelfef}
		&\EE_{P_{Y\hat YX_1X_2}}[\metric(X_2,Y) - \metric(X_1,Y)] \notag\\
		&= \sum_{\substack{P'_{\hat Y X}\in \Pc_n(\Xc\Yc): \\  |P'_{\hat Y X} - P_{\hat Y X}|_\infty \leq \frac{1}{n} }} \hspace{-1mm}\alpha(P'_{\hat Y X}) \EE_{P'_{Y\hat YX_1X_2}}[\metric(X_2,Y) - \metric(X_1,Y)] \\ \label{vnfceodedm}
		&< 0,
	\end{align}	
	where \eqref{fjegfmelfef} follows from \eqref{cnfralfefmefr} and \eqref{vnfceodedm} follows from \eqref{eq:maximaldefopt}. The above inequality contradicts the maximality assumption of $P_{Y\hat YX_1}$. Therefore, there must exist at least one $P'_{Y\hat Y X} $ in the sum \eqref{sgadfga} which is maximal. We call this maximal joint conditional distribution $\bar P_{Y\hat Y X}$.
	The distribution  $\bar P_{Y\hat Y X}$ is such that
	\begin{align}
		&\bar P_{X\hat Y} \in \Pc_n(\Xc\Yc) \\
		&|\bar P_{\hat Y X} - P_{\hat Y X}|_\infty \leq \frac{1}{n} \label{zfg09}\\
		&\bar P_{Y\hat Y X} = P_{Y|\hat Y X}\bar P_{\hat Y X}
	\end{align}
	fulfilling properties \eqref{eq:pbar1} and \eqref{eq:pbar2}. In addition we have that
	\begin{align}
		|\bar P_{XY}(j,k) &- P_{XY}(j,k)|\notag\\
		 &= \big|\sum_{k'} P_{Y|\hat Y X}(k|j,k') \bar P_{X\hat Y}(j,k') \\
		 & ~~~~~~~~~~~ -  P_{Y|\hat Y X}(k|j,k') P_{X\hat Y}(j,k') \big| \\
		&\leq  \sum_{k'}  |\bar P_{X\hat Y}(j,k') -  P_{X\hat Y}(j,k') | \label{asgffg345}\\
		&\leq \frac{K}{n}, \label{asgffg3457}
	\end{align}	
	where \eqref{asgffg345} follows from the triangle inequality and $P_{Y|\hat Y X}(k|j,k')\leq 1$ and \eqref{asgffg3457} follows from \eqref{zfg09}, proving property \eqref{eq:pbar3}.
	
	Now we have found a $\bar P_{Y\hat Y X}$  with properties \eqref{eq:pbar1}--\eqref{eq:pbar3}. We need to show that for this $\bar P_{Y\hat Y X}$, we have that
	\begin{align}\label{maintheoremeq345}
		\pemax(\Cc_n, \bar P_{Y|X}) \geq \gamma \petcemax(\Cc_n,\bar P_{\hat Y|X}).
	\end{align}
	In the following, we prove 	\eqref{maintheoremeq345}. Without loss of generality assume that $\xv_1$ is the codeword with maximum type conflict error on channel $P_{\hat Y|X}$. For every message $\ell=2\dotsc,M$, define the sets
	\begin{align}
		\Ac_{\ell} &= \{\yv \,|\, \metric^n(\xv_{\ell},\yv) \geq \metric^n(\xv_1,\yv) \}\\
		\Bc_{\ell} &= \{\hat \yv \,|\,\hat \pv_{\hat{\yv}|\xv_{\ell}} = \hat \pv_{\hat{\yv}|\xv_1} = \bar P_{\hat Y|X} \}.
	\end{align}
	The sets $\Ac_{\ell},\Bc_{\ell}$ are the sets of outputs that result in a pairwise mismatched decoding error or type-conflict error, respectively. Using these definitions we write the probability of mismatched decoding error over channel $\bar P_{Y|X}$ and the type-conflict error probability over channel $\bar P_{\hat Y|X}$ as
	\begin{align}
		&\pemax(\Cc_n,\bar P_{Y|X}) = \PP\bigg[\bigcup_{m'=2}^{M} \Ac_{m'} \bigg] \label{sfgjh}\\ \label{tcetce}
		&\petcemax(\Cc_n,\bar P_{\hat Y|X}) = \PP\bigg[\bigcup_{\ell=2}^{M} \Bc_{\ell}, \bigg]
	\end{align}	
	where both probabilities in \eqref{sfgjh} and \eqref{tcetce} are computed with respect to $\bar P_{Y|\hat YX}^n\times P_U$, where $P_U$ denotes the equiprobable distribution over the type class $\Tc_{\xv_1}(\bar P_{\hat Y|X})$.
	Also define 
	\beq
	\Dc_{\ell} = \Bc_{\ell} \backslash \cup_{i=1}^{\ell - 1} \Bc_{i}
	\eeq
	with $\Bc_0 = \emptyset$. Observe that while $\Bc_\ell$ are not necessarily disjoint, the newly constructed sets $\Dc_{\ell}$ are, and thus $\bigcup_{\ell=2}^{M} \Bc_{\ell} = \bigcup_{\ell=2}^{M} \Dc_{\ell}$. Then, we have
	\begin{align} \label{xhweindeidnme}
		\pemax(\Cc_n,\bar P_{Y|X}) 	 &= \PP\bigg[\bigcup_{m'=2}^{M} \Ac_{m'} \bigg] \\  \label{ceiferifmrrg}
		&\geq \PP\bigg[\bigcup_{m'=2}^{M} \Ac_{m'} \bigcap \bigcup_{\ell=2}^{M} \Bc_{\ell}\bigg]  \\ \label{cejfneifg444fm}
		&= \PP\bigg[\bigcup_{m'=2}^{M} \Ac_{m'} \bigcap \bigcup_{\ell=2}^{M} \Dc_{\ell}\bigg] \\ \label{veifnmerifnre}
		&= \sum_{\ell = 2}^{M} \PP\bigg[\bigcup_{m'=2}^{M} \Ac_{m'}| \Dc_{\ell}\bigg] \PP[\Dc_{\ell}] \\ \label{dvqwfbewufen}
		&\geq \sum_{\ell = 2}^{M} \PP[ \Ac_{\ell} | \Dc_{\ell} ] \PP[\Dc_{\ell}],
	\end{align}
	where \eqref{cejfneifg444fm} follows from $\bigcup_{\ell=2}^{M} \Bc_{\ell} = \bigcup_{\ell=2}^{M} \Dc_{\ell}$, \eqref{veifnmerifnre} follows from the fact that the sets $\Dc_\ell$ are disjoint and \eqref{dvqwfbewufen} is follows from lower bounding $\PP\bigg[\bigcup_{m'=2}^{M} \Ac_{m'}| \Dc_{\ell}\bigg]$ by $\PP[ \Ac_{\ell} | \Dc_{\ell} ]$. Although, inequality \eqref{dvqwfbewufen} has removed many error events, it does not weaken our bound since a type conflict error in the auxiliary channel induces a  $\metric$-decoding error in the original channel for the same codewords. 
	
	We now proceed to lower-bounding  $\PP[\Ac_{\ell} | \Dc_{\ell}]$. We first rewrite  $\PP[\Ac_{\ell} | \Dc_{\ell}]$ as follows
	\begin{align} \label{avdefnienfef}
		\PP[\Ac_{\ell} | \Dc_{\ell}] &= \PP[\metric^n(\xv_{\ell},\yv) \geq \metric^n(\xv_1,\yv)  | \Dc_{\ell}] \\
		&= \sum_{\hat \yv \in \Dc_\ell } \PP[\metric^n(\xv_{\ell},\yv) \geq \metric^n(\xv_1,\yv)  | \hat \yv ] \frac{\PP[\hat \yv]}{\PP[\Dc_\ell]}, \label{dfhdsf}
	\end{align}
	where the probability in \eqref{dfhdsf} is over output sequences $\yv$. A consequence of Lemma \ref{cbsidfeifnefi} is that, given that $\xv_1,\xv_\ell$ are fixed, $\PP[\metric^n(\xv_{\ell},\yv) \geq \metric^n(\xv_1,\yv)  |\hat\yv]$ depends on $\hat \yv$ only through their joint type, i.e.,
	\begin{align}
	\PP[\metric^n&(\xv_{\ell},\yv) \geq \metric^n(\xv_1,\yv)  | \hat \yv ]\notag\\
	& ~~=  \PP[\metric^n(\xv_{\ell},\yv) \geq \metric^n(\xv_1,\yv)  | \pv_{\hat Y X_1X_\ell} ],
	\label{adfhafdh}
	\end{align}
	where the joint type $\pv_{\hat Y X_1X_\ell}=\hat\pv_{\hat \yv\xv_1\xv_\ell}$.
	We now proceed to lower bound the right hand side of \eqref{adfhafdh} by using Lemma \ref{xvafjeofkeoe}.
	In order to apply the lemma to obtain a lower bound on $\PP[\metric^n(\xv_{\ell},\yv) - \metric^n(\xv_1,\yv) \geq 0 |\pv_{\hat Y X_1X_\ell} ]$ we proceed with the following steps:	
	\begin{enumerate}
		\item We derive a single-letter expression of the expectation $\EE[\metric^n(\xv_{\ell},\yv) -\metric^n(\xv_1,\yv)  |\pv_{\hat Y X_1X_\ell} ]$. To this end, we  use Lemma \ref{bfehdwbae}  for $Z_i=(\xv_1(i), \xv_\ell(i), \hat\yv(i))$ and $S_i=Y_i$ and $f(Z_i,S_i)=\metric(\xv_{\ell}(i), Y_i) - \metric(\xv_{1}(i), Y_i)$. Then, by using Lemma \ref{bfehdwbae} we obtain
		\begin{align}
			&\EE[\metric^n(\xv_{\ell},\yv) - \metric^n(\xv_1,\yv) | \pv_{\hat Y X_1 X_\ell }] \notag\\
			&~~=n \EE_{P_{Y|X_1,\hat Y} \times \pv_{\hat Y X_1 X_\ell }}[\metric(X_\ell,Y) - \metric(X_1,Y)]
		\end{align}
		As a result, since $\bar P_{Y\hat Y | X_1} \in \Mmax(q, \pv_X)$ is maximal, then 
		\begin{align}
			\EE_{P_{Y|X_1,\hat Y} \times \pv_{\hat Y X_1 X_\ell }}[\metric(X_\ell,Y) - \metric(X_1,Y) ] \geq 0.
		\end{align}
		\item We use  Corollary \ref{corollary1} to write the conditional  variance $\Var [\metric^n(\xv_{\ell},\yv) - \metric^n(\xv_{1},\yv) | \pv_{\hat Y X_1 X_\ell }] $ as
		\begin{align}
			&\Var [\metric^n(\xv_{\ell},\yv) - \metric^n(\xv_{1},\yv) | \pv_{\hat Y X_1 X_\ell }]\notag\\
			 &= n \EE_{ \pv_{\hat Y X_1 X_\ell }}\big[\Var_{P_{Y|\hat Y X_1}}[\metric(X_\ell,Y)- \metric(X_1,Y) ] \big].
		\end{align}
		\item From Lemma \ref{avgfemfaokeaa}, we have that $\metric^n(\xv_{\ell},\yv) - \metric^n(\xv_{1},\yv)$ given $\pv_{\hat Y X_1 X_\ell }$ is sub-Gaussian, i.e.,
		\begin{align}
			\PP[\metric^n(\xv_{\ell},\yv) - \metric^n(\xv_{1},\yv) \geq \xi| \pv_{\hat Y X_1 X_\ell }] \leq e^{\frac{-\xi^2}{n(b-a)^2} },
		\end{align}
		where $a = 2 \min_{x,y} \metric(x,y), b = 2 \max_{x,y} \metric(x,y)$.
		\item We apply Corollary \ref{cor:ineq_subg} to the random variable $\frac{\metric^{n}(\xv_{\ell},\yv)-\metric^{n}(\xv_{1},\yv)}{\sqrt{n}}$ and setting $\theta = \frac{1}{|a-b|}$ we obtain 
		\begin{align}\notag
			\PP[&\metric^{n}(\xv_{\ell},\yv)  -  \metric^{n}(\xv_1,\yv) \geq 0 | \pv_{\hat Y X_1 X_\ell } ] \notag\\
			&~~ \geq \frac{\EE_{ \pv_{\hat Y X_1 X_\ell }}\big[\Var_{P_{Y|\hat Y X_1}}[\metric(X_\ell,Y) - \metric(X_1,Y)]\big] }{2\kappa^2(a - b)^2} \notag \\
			&~~~~~~~~~~~~ - |a-b| e^{\frac{-\kappa^2}{2}} \Big( 1 + \sqrt{2} + \frac{\sqrt{2\pi}}{\kappa} + \frac{1}{\kappa^2}\Big).
			\label{sdhjsfgk}
		\end{align}
	\end{enumerate}
	
	The expected conditional variance in the right hand side of \eqref{sdhjsfgk} can potentially be very small. This can happen for types $ \pv_{\hat Y X_1 X_\ell }$ that have substantial mass in the entries where the conditional variance $\Var_{P_{Y|\hat Y X_1}}[\metric(X_\ell,Y) - \metric(X_1,Y)]$ is zero. This implies that conditioning on this type does not allow us to lower bound the probability by a constant, independent of $n$, as we would like.

	To overcome this problem, we shorten the code and received sequences by discarding the entries where the above conditional variance is zero. Then, we use again Corollary \ref{cor:ineq_subg}. More precisely,
	we define a new type $\tilde \pv_{\hat Y X_1 X_\ell }$ which places zero mass in the entries where the conditional variance
	\begin{align}
	&\sigma^2_{j_1,j_2,k}\triangleq \Var_{P_{Y|\hat Y X_1}}[\metric(X_\ell,Y) - \metric(X_1,Y) \notag\\
	&~~~~~~~~~~~~~~~~~~~~~~~~~~~~~~~| X_1 = j_1, X_\ell = j_2, \hat Y = k ]
	\label{eq:condvar}
	\end{align}
	 is zero
	\begin{align} \label{bcefneakfe}
		\pv^*_{\hat Y X_1 X_\ell} (k,j_1,j_2) = \begin{cases}
			0 \ \  \ &\sigma^2_{j_1,j_2,k} = 0\\
			\frac{ \pv_{\hat Y X_1 X_\ell }(k,j_1,j_2)}{ n^* } & \text{otherewise}
		\end{cases},
	\end{align}
	where $\tilde n\leq n$ is the length of the sequences after removing the zero-variance entries is defined as follows
	\begin{align} 
		n^* = n \sum_{j_1,j_2,k}  \pv_{\hat Y X_1 X_\ell }(k,j_1,j_2) \mathds{1}\big\{\sigma^2_{j_1,j_2,k}\neq 0\big\}.
	\end{align}
	This type consists of only the $k,j_1,j_2$ for which the conditional variance $\sigma^2_{j_1,j_2,k}$ in \eqref{eq:condvar} is $\sigma^2_{j_1,j_2,k} \neq 0$. We redefine the auxiliary channel output and the two codewords accordingly, by eliminating the entries with zero variance. More precisely, ${\xv^*_1},{\xv^*_\ell}, \hat \yv^*$ are defined by eliminating indices $0\leq i\leq n$ from $\xv_1,\xv_2,\hat\yv$ when $\xv_1(i)=j_1, \xv_\ell(i)=j_2, \hat \yv(i)=k $ and
	$\sigma^2_{j_1,j_2,k} = 0$.
	
	We define $\yv^*$ as the corresponding shortened length-$n^*$ channel output sequence. Then, we notice that 
	\begin{align} \label{bfvedfwfmefrt}
		&\PP[\metric^n(\xv_{\ell},\yv) \geq \metric^n(\xv_1,\yv) |  \pv_{\hat Y X_1 X_\ell } ] \notag\\
		&~~~~~~= \PP[\metric^{n^*}( \xv^*_{\ell}, \yv^*) \geq \metric^{n^*}( \xv^*_1, \yv^*)  + \mu^* |  \pv^*_{\hat Y X_1 X_\ell }],
	\end{align}
	where  we replace the zero-variance entries by
	\begin{align}
		\mu^* &=   \EE[\metric^{n^*}( \xv^*_{\ell}, \yv^*) - \metric^{n^*}( \xv^*_1, \yv^*) |  \pv^*_{\hat Y X_1 X_\ell }]\notag\\
		&~~~~~~~~~~~~~~~ -\EE[\metric^n(\xv_{\ell},\yv) - \metric^n(\xv_1,\yv) |  \pv_{\hat Y X_1 X_\ell } ] , 
	\end{align}
	where $\mu^*$ is the overall change in metric difference.

	Therefore, \eqref{bfvedfwfmefrt} follows from the fact that eliminating zero-variance entries at the positions as described in \eqref{bcefneakfe} corresponds to cases where the metric difference was a constant in that position.
	
	Notice that with the previous procedure we have
	\begin{align} \label{abdfeidemdefenfe}
		&\Var [\metric^{n^*}( \xv^*_{\ell}, \yv^*) - \metric^{n^*}( \xv^*_{1}, \yv^*) |  \pv^*_{\hat Y X_1 X_\ell }] \notag\\
		&= n^* \EE_{ \pv^*_{\hat Y X_1 X_\ell }}[\Var_{P_{Y|\hat Y X_1}}[\metric(X_\ell,Y)]- \metric(X_1,Y)] ] \\
		&\geq n^* \sigma^2,
	\end{align}
	where 
	\begin{align}
		\sigma^2 &= \min_{\substack{j_1,j_2,k :\\ \sigma^2_{j_1,j_2,k}> 0 }} \sigma^2_{j_1,j_2,k}\\
		&>0
	\end{align}	
	where $\sigma^2_{j_1,j_2,k}$ has been defined in \eqref{eq:condvar}.

	We now proceed to repeat steps 3) and 4) of the above procedure. We first use Lemma \ref{avgfemfaokeaa} and obtain that 
	\begin{align}	\label{cedfnmefkerf}
		\EE[\metric^{n^*}( \xv_{\ell}^*, \yv^*) - \metric^{n^*}( \xv_{1}^*, \yv^*) | \pv^*_{\hat Y X_1 X_\ell } ] = \mu & \geq \mu^*,
		\end{align} 
		\begin{align} \label{dfekrfermnere}
		\PP[ |\metric^{n^*}( \xv_{\ell}^*, \yv^*) - \metric^{n^*}( \xv_{1}^*, \yv^*) - \mu| &\geq \xi | \pv^*_{\hat Y X_1 X_\ell } ] \notag\\
		&\leq 2e^{\frac{-\xi^2}{n^*(b-a)^2} }.
	\end{align}
	
	We now apply Corollary \ref{cor:ineq_subg} as in step 4), and get that 
	\begin{align}
		&\PP[\metric^{n^*}( \xv_{\ell}^*, \yv^*) - \metric^{n^*}( \xv_{1}^*, \yv^*) \geq \mu^* | \pv^*_{\hat Y X_1 X_\ell} ] \notag\\
		& =\PP[\metric^{n^*}( \xv_{\ell}^*, \yv^*) - \metric^{n^*}( \xv_{1}^*, \yv^*) \geq \mu | \pv^*_{\hat Y X_1 X_\ell} ]\\
		&\geq \frac{\sigma^2}{2\kappa^2(a - b)^2} - |a-b| e^{\frac{-\kappa^2}{2}}\Big( 1 + \sqrt{2} + \frac{\sqrt{2\pi}}{\kappa} + \frac{1}{\kappa^2}\Big),
	\end{align}
	where $a = 2 \min_{x,y} \metric(x,y), b = 2 \max_{x,y} \metric(x,y)$.
	
	By setting $\kappa $ large enough we get a uniform bound for all $n^* > 0$. Let $\gamma > 0$ denote such a bound, i.e., 
	\begin{align} \label{gafdjeidfe}
		\PP[\metric^{n^*}( \xv_{\ell}^*, \yv^*) - \metric^{n^*}( \xv_{1}^*, \yv^*) \geq \mu^*|\pv^*_{\hat Y X_1 X_\ell}]  \geq \gamma
	\end{align}	
	for all $n^* > 0$. In case $n^* = 0$ the expression in left hand side of \eqref{gafdjeidfe} equals to 1 and the rest of the proof holds. Therefore, from \eqref{avdefnienfef} we get 
	\begin{align}
		\PP[\Ac_{\ell} | \Dc_{\ell}] &= \sum_{\hat \yv \in \Dc_\ell } \PP[\metric^n(\xv_{\ell},\yv) \geq \metric^n(\xv_1,\yv)  | \hat \yv ] \frac{\PP[\hat \yv]}{\PP[\Dc_\ell]} \\ \label{dedeifdfr}
		&\geq \gamma \sum_{\hat \yv \in \Dc_\ell } \frac{\PP[\hat \yv]}{\PP[\Dc_\ell]} \\ \label{cfaidjedje}
		& = \gamma,
	\end{align}	
	where \eqref{dedeifdfr} follows from \eqref{gafdjeidfe} and \eqref{cfaidjedje} follows from the fact that $\PP[\Dc_\ell] = \sum_{\hat \yv \in \Dc_\ell } \PP[\hat \yv]$.
	
	Therefore, combining the above inequality with \eqref{dvqwfbewufen} we get 
	\begin{align}
		\pemax(\Cc_n,\bar P_{Y|X}) &\geq \sum_{\ell = 2}^{M} \PP[ \Ac_{\ell} | \Dc_{\ell} ]\PP[\Dc_{\ell}] \\ \label{dveudneegr}
		&\geq \sum_{\ell = 2}^{M} \gamma \PP[\Dc_{\ell}] \\ \label{bvfvrjgrg}
		& = \gamma  \PP\big [\bigcup_{\ell = 2}^{M}\Dc_{\ell}\big]\\ \label{vijrginaacd}
		& = \gamma  \PP\big [\bigcup_{\ell = 2}^{M}\Bc_{\ell}\big] \\ \label{dedejfnergg}
		&= \gamma \petcemax(\Cc_n,\bar P_{\hat Y|X}),
	\end{align}
	where \eqref{dveudneegr} is deduced from \eqref{cfaidjedje}, \eqref{bvfvrjgrg} is resulted from the fact that the sets $\Dc_{\ell}$ are disjoint, \eqref{vijrginaacd}  follows from $\bigcup_{\ell=2}^{M} \Bc_{\ell} = \bigcup_{\ell=2}^{M} \Dc_{\ell}$ and \eqref{dedejfnergg}  follows from \eqref{tcetce}. This concludes the proof.

%

%

Unfortunately, the techniques introduced in the proof of Theorem \ref{maintheorem} do not seem to naturally extend to type-dependent metrics. This implies that the rate at which the error exponent derived in Section \ref{sec:tdm} becomes zero might not be the best possible bound to the mismatch capacity and might potentially be further improved, since there might be smaller rates where the error probability decays sub-exponentially.


\section{Auxiliary Lemmas} \label{useful_lemmas}
In this appendix we study expected values of functions under the equiprobable distribution over a type class. 
Let $\sv,\zv$ be sequences of length $n$ from alphabets $\Sc, \Zc$ respectively with joint type  $\hat \pv_{ \zv\sv} = \pv_{ZS}$.  Moreover, let $f: \Tc \times \Sc \to \RR$ be an arbitrary additive function, i.e., 
\begin{align}
f(\tv,\sv) = \sum_{i=1}^{n}f(t_i,s_i),
\end{align}
where with a slight abuse of notation we have used the same $f$ for sequences and their entries.

Let $P_U$ be the equiprobable distribution over all sequences $\tv$ such that $\hat \pv_{\tv \zv} = \pv_{TZ}$. In other words, $P_U$ denotes the equiprobable distribution over elements of the conditional type class $\Tc_{\zv}(\pv_{T|Z})$, where $\pv_{TZ}$ is a given type. The lemma below provides a single-letter expression for $\EE_{U}[f(\tv,\sv)]$.

\begin{lemma} \label{vrogjmfiarnqawen}
With the above assumptions we have 
\begin{align}
\EE_U[f(\tv,\sv)] = n\EE_{\pv_{T|Z} \times \pv_{ZS} }[f(T,S)].
\end{align}

\end{lemma}
\begin{IEEEproof}
We have
\begin{align}
\EE_{U}[f(\tv,\sv)] &= \EE_{U}\Big[\sum_{i=1}^{n}f(t_i,s_i)\Big]\\ \label{cvaufnafaf}
&= \sum_{i=1}^{n}\EE_{U}[f(t_i,s_i)] \\ \label{kveodadmangrta} 
& = \sum_{i=1}^{n}\EE_{\pv_{T|Z} \times \pv_{Z|S=s_i} }[f(T,S)|S= s_i] \\
& = \sum_{s}^{}n \pv_{S}(s)\EE_{\pv_{T|Z} \times \pv_{Z|S = s} }[f(T,S) | S = s] \\
& = n \EE_{\pv_{T|Z} \times \pv_{ZS} }[f(T,S)],
\end{align}
where \eqref{cvaufnafaf} follows from linearity of expectation and \eqref{kveodadmangrta} is deduced from $\hat \pv_{\tv \zv} = \pv_{TZ}$. 

\end{IEEEproof}
\begin{lemma} \label{xvafjeofkeoe}
	Let $Y$ be a zero-mean sub-Gaussian random variable with parameter $\theta$ i.e. $\PP[|Y| \geq \xi] \leq 2e^{\frac{-\xi^2 \theta^2}{2}}$ for all $\xi \geq 0$, then for any $a>0$ we have
	\begin{align}
		\PP[Y \geq 0] \geq \frac{\Var [Y]}{2a^2} - 2e^{\frac{-a^2\theta^2}{2}}\Big( 1 + \sqrt{2} + \frac{\sqrt{2\pi}}{a\theta} + \frac{1}{a^2\theta^2}\Big).
	\end{align}
\end{lemma}
\begin{IEEEproof}
	To begin with, we have that
	\begin{align} \label{jcodfmeofkeo}
		\mathds{1}\{Y \geq 0 \} \geq \frac{Y(Y+a)}{2a^2}\mathds{1}\{-a \leq Y \leq a \}.
	\end{align}
	For simplicity of notation let $\Ic = [-a,a]$. Therefore, by taking expectations from both sides of \eqref{jcodfmeofkeo} we get 
	\begin{align}
		\PP\{Y \geq 0 \} &\geq \EE\left[ \frac{Y(Y+a)\mathds{1}\{Y \in \Ic \}}{2a^2}\right] \\
		& = \EE\bigg[ \frac{Y^2\mathds{1}\{ Y \in \Ic \}}{2a^2} \bigg]+ \EE \bigg[ \frac{Y\mathds{1}\{ Y \in \Ic \}}{2a}\bigg] \\ 
		& = \EE\bigg[\frac{Y^2}{2a^2}\bigg] + \EE\bigg[\frac{Y}{2a}\bigg] - \EE\bigg[\frac{Y^2}{2a^2}\mathds{1}\{Y \notin \Ic \}\bigg] \notag\\
		&~~~~~~~~~~~~~~~~~~~~~~~~~ - \EE\bigg[\frac{Y}{2a} \mathds{1}\{Y \notin \Ic \} \bigg].
	\end{align}
	Now by substituting $\EE[Y] = 0$ and $\EE[Y^2] =  \Var [Y]$ we have
	\begin{align} \label{dvawfejafiwafrw}
		\PP\{Y \geq 0 \} \geq \frac{\Var[Y]}{2a^2} &- \EE\bigg[\frac{Y^2}{2a^2}\mathds{1}\{Y \notin \Ic \}\bigg] \notag\\
		&- \EE\bigg[\frac{Y}{2a} \mathds{1}\{Y \notin \Ic \} \bigg]
	\end{align}
	
	We now evaluate both expectations in \eqref{dvawfejafiwafrw}. We have that
	\begin{align} \label{cafjifmafgmf}
		\EE\bigg[\frac{Y^2}{2a^2}\mathds{1}\{Y \notin \Ic \}\bigg] &= \int_{0}^{\infty} \PP \Big( \frac{Y^2}{2a^2}\mathds{1}\{Y \notin \Ic \} > t \Big) dt \\ \label{cdkfeofkeo}
		& = \int_{0}^{\infty} \PP \Big( |Y| \geq \max\{\sqrt{2t}a,a\} \Big) dt \\ \label{ckrofkrfjwif}
		& \leq 2  \int_{0}^{\infty}  e^{ -\max\{2ta^2,a^2\} \theta^2 } dt \\ \label{vfjreifjaef}
		& = 2 \int_{0}^{ \frac{1}{2} }  e^{ -a^2 \theta^2 } dt + 2 \int_{\frac{1}{2}}^{ \infty }  e^{ -2ta^2 \theta^2 } dt\\ \label{ofrjfurehfewi}
		& = e^{ -a^2 \theta^2 }  + \frac{1}{a^2\theta^2} e^{ -a^2 \theta^2 },
	\end{align}  
	where \eqref{cafjifmafgmf} follows from rewriting the expectation, \eqref{ckrofkrfjwif} is followed from the sub-Gaussianity of $Y$.
	Similarly, we have
	\begin{align} 
		\EE\bigg[\frac{Y}{2a} &\mathds{1}\{Y \notin \Ic \}\bigg] \notag\\
		& \leq \EE\bigg[\frac{|Y|}{2a} \mathds{1}\{Y \notin \Ic \}\bigg] \\ \label{xvafiefnasrf}	
		&\leq \int_{0}^{\infty} \PP \Big( \frac{|Y|}{2a}\mathds{1}\{Y \notin \Ic \} > t \Big) dt \\ \label{djodmeodkew}
		& = \int_{0}^{\infty} \PP \Big( |Y| \geq \max\{2ta,a\} \Big) dt \\
		&\leq 2  \int_{0}^{\infty}  e^{ -\max\{2t^2a^2,a^2\} \theta^2 } dt \\
		& = 2 \int_{0}^{ \sqrt{\frac{1}{2}} }  e^{ -a^2 \theta^2 } dt + 2  \int_{ \sqrt{\frac{1}{2}} }^{ \infty }  e^{ -2t^2a^2 \theta^2 } dt \\ \label{hcwedojowJDW}
		&= \sqrt{2}  e^{ -a^2 \theta^2 } + 2\sqrt{2\pi}\frac{1}{2a\theta} Q(\sqrt{2}a\theta) \\
		& \leq  \sqrt{2}  e^{ -a^2 \theta^2 } +  \frac{\sqrt{2\pi}}{a\theta} e^{ -a^2 \theta^2 }, \label{xvafueafne}
	\end{align}
	where $Q$ is the Gaussian $Q$-function, \eqref{xvafiefnasrf} follows from the sub-Gaussianity of $Y$ and \eqref{hcwedojowJDW} follows from the change of variable $u = 2a\theta t$. Moreover,  \eqref{xvafueafne} is resulted from using the Chernoff bound on the $Q$-function.
	by substituting \eqref{ofrjfurehfewi} and \eqref{xvafueafne} in \eqref{dvawfejafiwafrw} we get the desired result.
	
\end{IEEEproof}
\begin{corollary}\label{cor:ineq_subg}
	For a sub-Gaussian random variable $Z$ with parameter $\theta$ i.e. $\PP[|Z - \EE[Z]| \geq \xi] \leq 2e^{\frac{-\xi^2 \theta^2}{2}}$ for all $\xi \geq 0$, for any $\kappa > 0$ we get
	\begin{align}
		\PP[Z \geq \EE[Z]] \geq \frac{\theta^2 \Var [Z]}{2\kappa^2} - 2e^{\frac{-\kappa^2}{2}} \Big( 1 + \sqrt{2} + \frac{\sqrt{2\pi}}{\kappa} + \frac{1}{\kappa^2}\Big).
	\end{align}
\end{corollary}
\begin{IEEEproof}
	By setting $a = \frac{\kappa}{\theta}$ and substituting $Y = Z - \EE[Z]$ in the above lemma we get the above inequality. 
\end{IEEEproof}
We will use this form of the inequality throughout the paper.

The next lemma compares the size of the type class $\Tc(\pv_{ZSU})$ for joint type $\pv_{ZSU}$,  with number of sequences whose marginal types are equal to $\pv_{ZS}$ and $\pv_{SU}$.

\begin{lemma} \label{oefkznfdwarnawe}
Consider type $\pv_{ZSU}$ which is the multiplication of two types $\pv_{ZSU} = \pv_{Z|S}\pv_{SU}$. We have the following inequality
\begin{align}
\frac{\big|\{ (\zv,\sv,\uv) \in \Zc \times \Sc \times \Uc| \hat \pv_{\zv \sv \uv } = \pv_{ZSU} \} \big|}{\big|\{ (\zv,\sv,\uv) \in \Zc \times \Sc \times \Uc| \hat \pv_{\zv \sv } = \pv_{ZS}, \hat \pv_{\sv \uv } = \pv_{SU} \}\big|} \geq 2^{-n\omega_n},
\end{align}      
where $\omega_n = \Oc(\frac{\log n}{n})$.
\end{lemma}

\begin{IEEEproof}
From method of types properties (see e.g. \cite{gallager_fcc_notes, csiszar2011information}) we have
\begin{align} \label{rbsfneihgmtgir}
\big|\{ (\zv,\sv,\uv) \in \Zc \times \Sc \times &\Uc| \hat \pv_{\zv \sv \uv } = \pv_{ZSU} \} \big|\notag\\
 &= 2^{n(H(Z,S,U)+\zeta_n)} \\ \label{bfweodkaefeofef}
&= 2^{n(H(Z|S,U) + H(S,U) + \zeta_n)} \\ \label{bdqfmwergtmtor}
& = 2^{n(H(Z|S) + H(S,U) + \zeta_n)},
\end{align}
where $\zeta_n = \Oc(\frac{\log n}{n})$, the entropies in the above expressions are computed with respect to probability distribution 
$\pv_{ZSU}$, \eqref{rbsfneihgmtgir} follows from counting elements of a type class \cite{gallager_fcc_notes}, \eqref{bfweodkaefeofef} is derived by using the chain rule of entropy and \eqref{bdqfmwergtmtor} is deduced by using $\pv_{ZSU} = \pv_{Z|S}\pv_{SU}$ implying that $Z$ is independent of $U$ given $S$.

On the other hand 
\begin{align} 
&\big|\{ (\zv,\sv,\uv) \in \Zc \times \Sc \times \Uc| \hat \pv_{\zv \sv } = \pv_{ZS}, \hat \pv_{\sv \uv } = \pv_{SU} \}\big| \\ \label{aoefeifnefnek}
&= \sum_{  \substack{\tilde \pv_{ZSU} \\  \hat \pv_{\zv \sv } = \tilde \pv_{ZS}, \\\hat \pv_{\sv \uv}  = \tilde \pv_{SU}}} \big|\{ (\zv,\sv,\uv) \in \Zc \times \Sc \times \Uc| \hat \pv_{\zv \sv \uv } = \tilde \pv_{ZSU} \}\big| \\ \label{dndfoafpepf}
&\leq (n+1)^{|\Zc||\Sc||\Uc|} \notag\\
&\max_{  \substack{\tilde \pv_{ZSU} \\  \hat \pv_{\zv \sv } = \tilde \pv_{ZS},\\ \hat \pv_{\sv \uv}  = \tilde \pv_{SU}}} \big|\{ (\zv,\sv,\uv) \in \Zc \times \Sc \times \Uc| \hat \pv_{\zv \sv \uv } = \tilde \pv_{ZSU} \}\big|   \\ \label{adpwkfpgepe}
&= 2^{n(H(Z,S,U)+\theta_n)} \\ \label{scfneifbrokgro}
& = 2^{n(H(Z|S,U) + H(S,U) +\theta_n)} \\ \label{btpdmanfeam}
&\leq 2^{n(H(Z|S) + H(S,U) +\theta_n)},
\end{align}
where $\theta_n = \Oc(\frac{\log n}{n})$, \eqref{aoefeifnefnek} is derived by considering all types with marginals $\pv_{ZS}$ and $\pv_{SU}$, \eqref{dndfoafpepf} follows by upper bounding the number of types with $(n+1)^{|\Zc||\Sc||\Uc|}$ and number of elements of each type class with the number of elements of the largest one, and \eqref{adpwkfpgepe} follows by counting the elements of the type class that maximises the expression 
\begin{align}
\max_{  \substack{\tilde \pv_{ZSU} \\  \hat \pv_{\zv \sv } = \tilde \pv_{ZS},\\ \hat \pv_{\sv \uv}  = \tilde \pv_{SU}}} \big|\{ (\zv,\sv,\uv) \in \Zc \times \Sc \times \Uc| \hat \pv_{\zv \sv \uv } = \tilde \pv_{ZSU} \}\big|.
\end{align} 
In other words, the entropy in \eqref{adpwkfpgepe} is computed with respect to this maximising type $\tilde \pv_{ZSU}$ in the previous expression. In the proceeding expressions, \eqref{adpwkfpgepe},  \eqref{scfneifbrokgro} and \eqref{btpdmanfeam} the same type and its corresponding marginals are used. Eq. \eqref{scfneifbrokgro} is derived using chain rule of entropy and \eqref{btpdmanfeam} follows from $H(Z|S,U) \leq H(Z|S)$. Since the marginals of both types $\pv_{ZSU}$ and $\tilde\pv_{ZSU}$ are the same, $H(Z|S), H(S,U)$ are the same in \eqref{btpdmanfeam} and \eqref{bdqfmwergtmtor}. Therefore, we have that
\begin{align}
&\frac{\big|\{ (\zv,\sv,\uv) \in \Zc \times \Sc \times \Uc| \hat \pv_{\zv \sv \uv } = \pv_{ZSU} \} \big|}{\big|\{ (\zv,\sv,\uv) \in \Zc \times \Sc \times \Uc| \hat \pv_{\zv \sv } = \pv_{ZS}, \hat \pv_{\sv \uv } = \pv_{SU} \}\big|}\notag\\
&~~~~~~~~~~~~~~~~~~~~~~~~~~~~~~~~~~~~~~~~~~  \geq 2^{n(\zeta_n - \theta_n)}.
\end{align}
By setting $\omega_n = \zeta_n - \theta_n$ we get the desired result. 
\end{IEEEproof}

Before stating the next lemma we need to define the convex hull of a set of vectors. 
\begin{definition}
	Let $\vv_1,\cdots, \vv_\ell\in \RR^d$ be vectors in a vector space. Then convex hull of these vectors denoted by $\text{\rm CVH}(\{\vv_1,\cdots,\vv_\ell\})$ is defined as the following set
	 \begin{align}
	 	&\text{\rm CVH}(\{\vv_1,\cdots,\vv_\ell\}) \notag\\
		&= \bigg \{\vv\in\RR^d~\big| ~\vv = \sum_{i=1}^{\ell} \alpha_i \vv_i, \alpha_i\in[0,1],\sum_{i=1}^{\ell} \alpha_i = 1 \bigg \}.
	 \end{align}
\end{definition}

\begin{lemma} \label{bogjamirenruanr}
	Let $\pv_{ZS},\pv_{SU}$ be two joint types. Define the distribution $P_{ZSU}^\star = \pv_{Z|S}\pv_{SU}$ and let 
	\begin{align}
		&\Ac = \Big \{\pv_{Z_1S_1U_1 } \in \Pc_n(\Zc \times \Sc \times \Uc) | \pv_{Z_1S_1} = \pv_{ZS}, \notag\\
		&~~~~~~~~~~~~\pv_{S_1U_1} = \pv_{SU}, |\pv_{Z_1S_1U_1 } - P_{ZSU}^\star|_{\infty} \leq \frac{1}{n} \Big \}. 
	\end{align}
Then $P_{ZSU}^\star \in \text{\rm CVH}(\Ac)$.	
\end{lemma}

The implication of the above lemma is that while $P_{ZSU}^\star$ is not necessarily a type it can be expressed as a convex combination of types that are in the neighborhood of $ P_{ZSU}^\star$ and also has marginals equal to the marginals of $P_{ZSU}^\star$, which are types by definition.
\begin{IEEEproof}
	We prove a stronger result than the one mentioned in the theorem statement. We have used a simpler version of the lemma's statement since it will be what we need in the proofs of the main results of this paper. We will use induction to prove the the following statement.
    For every probability distribution $P_{ZSU}^\star$ and set $\Jc \subseteq \Zc \times \Sc \times \Uc$, we have $P^\star_{ZSU}\in\text{\rm CVH}(\Ac_{\Jc})$, where
\begin{align}
&\Ac_{\Jc} = \Big \{\pv_{Z_1S_1U_1 } \in \Pc_n(\Zc \times \Sc \times \Uc) ~|~ \pv_{Z_1S_1} = \pv_{ZS}, \notag\\
&~~~~~~~~\pv_{S_1U_1} = \pv_{SU}, |\pv_{Z_1S_1U_1 } - P_{ZSU}^\star|_{\infty} \leq \frac{1}{n} , \notag\\
 &~~~~~~~~\forall (i,j,k) \in \Jc, \pv_{Z_1S_1U_1 }(i,j,k) = P_{ZSU}^\star(i,j,k) \Big \}.
\end{align}   
Define the hyperplane $\Hc$ as the set
	\begin{align} \label{xbaeofmefoe}
		\Hc = \{  P_{Z_1 S_1U_1} |  P_{Z_1S_1} = \pv_{ZS}, P_{S_1U_1} = \pv_{SU} \}
	\end{align}
where in the definition of $\Hc$, the quantities $P_{Z_1 S_1U_1}$ are not assumed to satisfy $P_{Z_1S_1U_1}(i,j,k) \geq 0$, but they satisfy $\sum_{z,s,u} P_{Z_1 S_1U_1}(z,s,u)=1$. This makes the above set a hyperplane. Therefore, $P_{ZSU}^\star\in \Hc$, but since it is a distribution, it satisfies that $P_{ZSU}^\star(i,j,k) \geq 0$. Define also the set 
\begin{align} \label{xbaeofmefoej}
		&\Hc_\Jc = \{  P_{Z_1 S_1U_1} | \forall (i,j,k) \in \Jc, P_{Z_1S_1}(i,j) = \pv_{ZS}(i,j), \notag\\
		&~~~~~~~~~~~~~~~P_{S_1U_1}(j,k) = \pv_{SU}(j,k) \} \subseteq \Hc.
	\end{align}

We perform the induction on the dimension, or number of degrees of freedom, of the set $\Hc_\Jc$. Recall that $P_{ZSU}^\star$ is not necessarily a type but its marginals $P_{ZS}^\star=\pv_{ZS}, P_{SU}^\star=\pv_{SU}$ are types.

Additionally, define the set $\Bc$
\begin{align}
	\Bc &= \Big\{ \pv_{Z_1S_1U_1}~ |~ \forall (i,j,k) ~\pv_{Z_1S_1U_1}(i,j,k) \geq 0,\notag\\
	 &~~~~|\pv_{Z_1S_1U_1 } - P_{ZSU}^\star|_{\infty} \leq \frac{1}{n} \notag\\ 
	&~~~~\forall (i,j,k) \in \Jc, \pv_{Z_1S_1U_1 }(i,j,k) = P_{ZSU}^\star(i,j,k) \Big\},
\end{align}
where in the definition of the set $\Bc$, the quantities $\pv_{Z_1S_1U_1}$ are not assumed to sum to $1$, but instead, they are assumed to satisfy $\pv_{Z_1S_1U_1}(i,j,k) \geq 0$ for all $(i,j,k)$.

Then, we deduce that $P_{ZSU}^\star \in \text{\rm CVH} (\Bc) \cap \Hc$ because $P_{ZSU}^\star$ belongs to both $\text{\rm CVH} (\Bc )$ and $\Hc$.
Moreover, the intersection of $\text{\rm CVH} (\Bc )$ and $\Hc$ is a convex set since the intersection of any convex set and a hyperplane is a convex set. 

For any $\tilde \Jc \supset \Jc$ and $P_{ZSU}^\dagger$ where $\forall (i,j,k) \in \Jc$ we have $P_{ZSU}^\dagger(i,j,k) = P_{ZSU}^\star(i,j,k)$ we define a side of set $\text{\rm CVH} (\Bc)$ as the set $\text{\rm CVH} (\Bc_{\tilde \Jc})$, where
\begin{align}
	\Bc_{\tilde \Jc} &= \Big\{ \pv_{Z_1S_1U_1}~ |~ \forall (i,j,k), ~\pv_{Z_1S_1U_1}(i,j,k) \geq 0,\notag\\
	 &~~~~|\pv_{Z_1S_1U_1 } - P_{ZSU}^\star|_{\infty} \leq \frac{1}{n} \notag\\ 
	&~~~~\forall (i,j,k) \in \tilde \Jc, \pv_{Z_1S_1U_1 }(i,j,k) = P_{ZSU}^\dagger(i,j,k) \Big\}
\end{align}
We claim that the intersection of any side of $\text{\rm CVH} (\Bc)$ with set $\Hc_{\Jc}$, i.e.,  $\text{\rm CVH} (\Bc_{\tilde \Jc}) \cap \Hc_{\Jc}$ is the convex hull of all types in $\text{\rm CVH} (\Bc_{\tilde \Jc}) \cap \Hc_{\Jc}$.

Observe that if we prove this, then the induction step is proved. This is true since if the previous claim is proven, we would deduce that $\text{\rm CVH} (\Bc ) \cap \Hc_{\Jc}$ is itself the convex hull of types in $\text{\rm CVH} (\Bc ) \cap \Hc_{\Jc}$. As a result, any element of $\text{\rm CVH} (\Bc ) \cap \Hc_{\Jc}$ including $P_{ZSU}^\star$ can be expressed as a convex combination of types in $\text{\rm CVH} (\Bc ) \cap \Hc_{\Jc}$. Additionally, observe that from their definition, the set of types that belong to $\text{\rm CVH} (\Bc ) \cap \Hc_{\Jc}$ is equal to $\Ac_{\Jc}$.

To prove our claim we notice that for any $\Jc \subset \tilde \Jc$ we have that
\begin{align}
	\text{\rm CVH} (\Bc_{\tilde \Jc}) \cap \Hc_{\Jc} = \text{\rm CVH} (\Bc_{\tilde \Jc}) \cap \Hc_{\tilde \Jc}.
\end{align}
In addition, observe that
\begin{align}
	\text{\rm CVH} (\Bc_{\tilde \Jc}) \cap \Hc_{\tilde \Jc} = \Ac_{\tilde \Jc}.
\end{align}

Therefore, from the induction step we deduce that any distribution $\tilde P_{ZSU} \in \text{\rm CVH} (\Bc_{\tilde \Jc}) \cap \Hc_{\tilde \Jc}$ can be expressed as a convex combination of types in this side $\text{\rm CVH} (\Bc_{\tilde \Jc}) \cap \Hc_{\tilde \Jc}$. Therefore, the desired induction step is proven. 
Since we perform induction over the dimension of $\Hc_\Jc$, when this dimension is $1$, this is a trivial statement. Therefore, the proof by induction is complete.
\end{IEEEproof}

\begin{lemma} \label{odeormewr}
	Let $\pv_{X}$ be a type and $P_X$ be a distribution such that $|\pv_{X} - P_X|_{\infty} \leq \frac{1}{n}$. Then we have 
	\begin{align}
		&|\Tc(\pv_X)| = 2^{nH(X) - \omega_n} \\
		& |\omega_n| \leq \frac{|\Xc|\log n }{n},
	\end{align}
	where the entropy is computed with respect to distribution $P_X$.
\end{lemma}
\begin{IEEEproof}
	This lemma has been proven in \cite{gallager_fcc_notes}.
\end{IEEEproof}
The following lemma is eventually used for connecting type conflict errors of the auxiliary channel $P_{\hat Y|X}$ and mismatch decoding errors of $P_{Y|X}$. Since for such connections we use a bipartite graph and not all of the edges of the bipartite graph are useful we need the following lemma as a lower bound to the number of erroneous sequences under mismatched decoding.
\begin{lemma} \label{graphlemma}
	Let $\Gc_{\xv}(\pv_{Y\hat Y |X})$ be a regular bipartite graph between type classes $\Tc_{\xv}(\pv_{Y|X})$ and $\Tc_{\xv}(\pv_{\hat Y|X})$ with right degree $r_1$ and left degree $r_2$. Construct a graph $\tilde \Gc_{\xv}(\pv_{Y'\hat{Y}|X_1})$ by removing  some edges connected to  every $\hat \yv \in\Tc_{\xv}(\pv_{\hat Y|X})$ in such a way that at least $\alpha r_1$ of these edges for $ 0< \alpha < 1$ remain.
Then, for every set $\Bc \subset \Tc_{\xv}(\pv_{\hat Y|X})$ we have
	
	\begin{align}
		\frac{|\Psi(\Bc)|}{|\Tc_{\xv}(\pv_{Y|X})|} \geq \alpha  \frac{|\Bc|}{|\Tc_{\xv}(\pv_{\hat Y|X})|},
	\end{align}
where $\Psi(\Bc)$ is defined as
\begin{align}
\Psi(\Bc)&= \big\{\yv \in \Tc_{\xv_1}(\pv_{Y|X_1}) \ |\ \yv \text{ is connected}\notag\\
&~~~~~~~~~\text{ to some } \hat \yv \in \Bc \text{ in graph } \tilde \Gc_{\xv}(\pv_{Y\hat{Y}|X_1}) \big\}.
\end{align}
\end{lemma}

\begin{IEEEproof}
Observe that when we eliminate some edges, the degree of every $\yv \in \Tc_{\xv}(\pv_{Y|X})$ is at most $r_2$ and degree of every element in $\hat \yv \in \Tc_{\xv}(\pv_{\hat Y|X})$ is at least $\alpha r_1$. Therefore, if we count the number of edges between $\Bc$ and $\Psi(\Bc)$ and call it $e$, we have
\begin{align} \label{vjsfioamrawr}
 \alpha|\Bc| r_1 \leq e \leq r_2 |\Psi(\Bc)|.
\end{align}
The above inequality holds since degree of every vertex in $\Bc \subset \Tc_{\xv}(\pv_{\hat Y|X})$ is at least $\alpha r_1$ therefore, $e$ is at least $|\Bc| \alpha r_1$. On the other hand, degree of every element in $\Psi(\Bc)$ is at most $r_2$. As a result, $e$ is at most $|\Psi(\Bc)|r_2$.
In addition, observe that
\begin{align} \label{abfekfefd}
|\Tc_{\xv}(\pv_{\hat Y|X})| r_1 = |\Tc_{\xv}(\pv_{ Y|X})|r_2.
\end{align}
which follows by counting the number all edges in graph $\Gc_{\xv}(\pv_{Y\hat Y |X})$.
As a result, by substituting $\frac{r_1}{r_2}$ from \eqref{abfekfefd} in \eqref{vjsfioamrawr} we get
\begin{align}
\frac{|\Psi(\Bc)|}{|\Tc_{\xv}(\pv_{Y|X})|} \geq \alpha  \frac{|\Bc|}{|\Tc_{\xv}(\pv_{\hat Y|X})|}.
\end{align}
     
\end{IEEEproof}

\begin{lemma} \label{bcedfkmelfmefe}
	Let $P_{Y X}, \bar P_{Y X}$ be two joint distributions such that $|\bar P_{XY} - P_{XY}  |_{\infty} \leq \frac{K}{n}$. Then, there exists an $N_0$ such that for any $n>N_0$ and for any pair of sequences $(\xv,\yv)\in\Xc^n\times\Yc^n$ with joint type $\hat \pv_{\xv\yv}$ we have
	\begin{align}
		 e^{-\delta} \leq \prod_{i=1}^n \frac{P_{XY}(x_i,y_i)}{\bar P_{XY}(x_i,y_i)}\leq e^\delta,
	\end{align}
	where $\delta = \frac{2K }{\min_{P_{XY}(j,k)>0} P_{XY}(j,k)}$.
	
	In addition, for any codebook $\Cc_n$
	\begin{align}
	\pemax(\Cc_n, P_{Y|X}) \geq e^{-\delta} \pemax(\Cc_n, \bar P_{Y|X}).
	\end{align}
\end{lemma}
\begin{IEEEproof}
     We have the following 
     \begin{align}
     	 &\prod_{i=1}^n \frac{P_{XY}(x_i,y_i)}{\bar P_{XY}(x_i,y_i)}\notag\\
	  &= \prod_{j,k} \left[ \frac{P_{XY}(j,k) }{ \bar P_{XY}(j,k) }  \right]^{n\hat \pv_{\xv\yv}(j,k)}  \\
     	 &= \prod_{j,k} \left[ \frac{\bar P_{XY}(j,k) + P_{X,Y}(j,k) - \bar P_{X,Y}(j,k)  }{ \bar P_{XY}(j,k) }  \right]^{n\hat \pv_{\xv\yv}(j,k)} \\
     	 &\leq \prod_{j,k} \left[ \frac{\bar P_{XY}(j,k) + |P_{XY}(j,k) - \bar P_{XY}(j,k)|  }{ \bar P_{Y|X}(j,k) }  \right]^{n\hat \pv_{\xv\yv}(j,k)} \\ \label{rfngrkgerklg}
     	 &\leq \prod_{j,k}\Big(1+  \frac{\delta}{n} \Big)^{n\hat \pv_{\xv\yv}(j,k)} \\
     	 &= \Big(1+ \frac{\delta}{n}\Big)^n \\ \label{defjefmeme}
     	 & \leq e^\delta,
     \end{align}
     where \eqref{rfngrkgerklg} follows from $|\bar P_{XY} - P_{XY}  |_{\infty} \leq \frac{K}{n}$ and the definition of $\delta$. Moreover, there exists an $N_0$, such that for $n>N_0$ we have that $\bar P_{Y|X}(j,k) \geq \frac{P_{Y|X}(j,k)}{2}$. 
          The other inequality is derived similarly.
 	 
 	As a result, without loss of generality assume $\xv_1$ is the codeword with maximum probability of error and $\Bc$ be the set of all output sequences such that cause a $\metric$-decoding error when $\xv_1$ is sent. Therefore,
 	\begin{align}
 		\pemax(\Cc_n, P_{Y|X}) &= \PP[\Bc|\xv_1] \label{asfgsfg}\\
 		&\leq e^{\delta} \bar \PP[\Bc|\xv_1] \label{asfgsfg2}\\
 		&= e^{\delta}\pemax(\Cc_n, \bar P_{Y|X}),
 	\end{align}
 	where the probabilities in \eqref{asfgsfg} and \eqref{asfgsfg2} are computed with respect to $P_{XY}$ and $\bar P_{XY}$, respectively. This concludes the proof.
\end{IEEEproof}

\section{Conditioning on the type of a sequence } \label{xvajfefodefmer}

In this section, we study the effect of conditioning on the type of a sequence when computing some statistical properties of functions of random sequences. 

	\begin{lemma} \label{cbsidfeifnefi}
		Let $f: \Zc \times \Sc \to \RR$ be an arbitrary function and $(Z_i,S_i), i = 1,2,\dotsc,n$ be random variables taking values on alphabets $\Zc, \Sc $, respectively. Further assume that 
			\begin{align}\label{eodeodenfbfnvb}
				\PP \Big[S = \sv  \Big | Z = \zv\Big] = \prod_{i = 1}^{n}\PP [S_i = s_i | Z_i = z_i]
			\end{align}
			and $P_{Z_iS_i}$ does not depend on index $i$. Let $\hat \pv_\zv, \hat \pv_\sv$ denote the types of $\zv = (z_1,\dotsc,z_n),\sv = (s_1,\dotsc,s_n)$, respectively. Then, for any function $g$ the expectation 
		\beq
		\EE\bigg[g\Big(\sum_{i = 1}^{n} f(Z_i,S_i)\Big) \Big | Z = \zv \bigg]
		\eeq
		only depends on $\hat \pv_\zv$.
	\end{lemma}
	\begin{IEEEproof}
		It is sufficient to show that for any $\zv_1,\zv_2$ that $\hat \pv_{\zv_1} = \hat \pv_{\zv_2}$ we have 
		\begin{align}
			&\EE\bigg[g\Big(\sum_{i = 1}^{n} f(Z_i,S_i)\Big) \Big | Z = \zv_1 \bigg] \notag\\
			&~~~~~~~~~~~~~~~~= \EE\bigg[g\Big(\sum_{i = 1}^{n} f(Z_i,S_i)\Big) \Big | Z = \zv_2 \bigg].
		\end{align}
		We have 
			\begin{align}
				&\EE\bigg[g\Big(\sum_{i = 1}^{n} f(Z_i,S_i)\Big) \Big | Z = \zv \bigg] \notag\\
				&= \int_{\Sc^n}^{}g\Big(\sum_{i = 1}^{n} f(z_i,s_i)\Big) \PP \Big[S_1 = s_1,\cdots, S_n=s_n  \Big | Z = \zv\Big]dS \\ \label{freifeif}
				&= \int_{\Sc^n}^{}g\Big(\sum_{i = 1}^{n} f(z_i,s_i)\Big) \prod_{i = 1}^{n}\PP [S_i = s_i | Z_i = z_i]dS
			\end{align}
			where \eqref{freifeif} follows from \eqref{eodeodenfbfnvb}. Now notice that with a permutation of indices we can turn $\zv_1$ into $\zv_2$. Moreover, the expression in \eqref{freifeif} is invariant under permutation of indices because $P_{Z_iS_i}$ does not depend on index $i$. Therefore, the expression in \eqref{freifeif} is equal for $\zv_1$ and $\zv_2$. This finishes the proof.
		
	\end{IEEEproof}

Having the above lemma in mind, we study the problem of conditioning on types in the following results.
\begin{lemma} \label{bfehdwbae}
	Under the assumptions of Lemma \ref{cbsidfeifnefi}, we have that
	\begin{align}
		\EE\left[\sum_{i = 1}^{n} f(Z_i,S_i) \Big | \hat \pv_\zv \right] = n \EE_{P_{S|Z} \times \hat \pv_\zv}[f(\tilde Z,S)] 
	\end{align}
	where $\tilde Z$ is a random variable with distribution $\hat \pv_\zv $. 
\end{lemma}
\begin{IEEEproof}
We have that
	\begin{align} \label{cdbyfegf}
		\EE\left[\sum_{i = 1}^{n} f(Z_i,S_i) \Big |\hat \pv_\zv \right] &= \EE\left[\sum_{i = 1}^{n}\sum_{z}^{}f(z,S_i)\hat \pv_\zv(z)\right]  \\ \label{datwvfef}
		&= \sum_{i=1}^{n}\sum_{z}^{} \EE \left[ f(z,S_i)\right] \hat \pv_\zv(z) \\ \label{ofaeiufhes}
		&= \sum_{i=1}^{n}\sum_{z}^{} \EE_{P_{S|\tilde Z = z}} [f(z,S)] \hat \pv_\zv(z) \\
		&= n\EE_{\hat \pv_\zv}\left[\EE_{P_{S|\tilde Z}} \left[f(\tilde Z,S)\Big| \tilde Z \right] \right]\\  \label{ofvrsgbeg}
		& = n \EE_{P_{S|\tilde Z} \times \hat\pv_\zv(z)}[f(\tilde Z,S)] ,
	\end{align}
	where \eqref{cdbyfegf} is derived using the fact that $P_{Z_iS_i}$ does not depend on index $i$, \eqref{datwvfef} is derived from the linearity of expectation, \eqref{ofaeiufhes} follows by replacing random variables $S_i$ by $S$ which does not affect the expectation and \eqref{ofvrsgbeg} follows from the tower rule of conditional expectation. 
\end{IEEEproof}
\begin{lemma}  \label{lemma2}
	Under the assumptions of Lemma \ref{bfehdwbae} we have
	\begin{align}
		&\EE \bigg[\bigg(\sum_{i=1}^{n} f(Z_i,S_i)\bigg)^2 \Big| \hat \pv_\zv\bigg] = n^2 \EE_{P_{S|Z} \times \hat \pv_\zv}[f(\tilde Z,S)] ^2\notag\\
		& ~~~~~~+  n \EE_{P_{S|Z} \times \hat \pv_\zv}[f(\tilde Z,S)^2] - n\EE_{\hat \pv_\zv} \left[ \EE_{P_{S|Z}} [f(\tilde Z,S)|\tilde Z]^2\right],
	\end{align}
	where $\tilde Z$ is a random variable with distribution $\hat \pv_\zv $.
\end{lemma}
\begin{IEEEproof}
	By expanding the term in the expectation we have
	\begin{align} \label{ptgfaefqw}
		&\EE \bigg[\bigg(\sum_{i=1}^{n} f(Z_i,S_i)\bigg)^2 \Big| \hat \pv_\zv\bigg]\notag\\
		 &= 
		\EE\bigg[\sum_{i\neq k}^{} f(Z_i,S_i)f(Z_k,S_k) \Big| \hat \pv_\zv\bigg] +\EE\bigg[\sum_{i=1}^{n} f(Z_i,S_i)^2 \Big| \hat \pv_\zv\bigg]
	\end{align}
	Then for the first term of the right hand side of \eqref{ptgfaefqw} we can use Lemma \ref{bfehdwbae}
	\begin{align}\label{firstterm}
		\EE\bigg[\sum_{i=1}^{n} f(Z_i,S_i)^2 \Big| \hat \pv_\zv\bigg] = n \EE_{P_{Z|S} \times \hat \pv_\zv}[f(\tilde Z,S)^2],
	\end{align}
	where $\tilde Z$ is a random variable with distribution $\hat \pv_\zv $. Moreover, for the second term of right hand side of  \eqref{ptgfaefqw} we have
	\begin{align}
		&\EE\bigg[\sum_{i\neq k}^{} f(Z_i,S_i)f(Z_k,S_k) \Big| \hat \pv_\zv \bigg] \notag\\
		&= \EE \bigg[ \sum_{z_1 \neq z_2} \sum_{i\neq k} f(z_1,S_i)f(z_2,S_k) \hat \pv_\zv(z_1) \frac{n \hat \pv_\zv(z_2)}{n-1} \bigg] \notag\\
		&~~~~+ \EE \bigg[ \sum_{z}\sum_{i\neq k} f(z,S_i)f(z,S_k)\hat \pv_\zv(z) \frac{n\hat \pv_\zv(z) - 1}{n-1} \bigg]  \label{gusuerat}\\ 
		&= \frac{n}{n-1}\EE \bigg[ \sum_{i \neq k} \sum_{z_1,z_2} f(z_1,S_i)\hat \pv_\zv(z_1) f(z_2,S_k) \hat \pv_\zv(z_2) \bigg] \notag\\
		& ~~~~- \frac{1}{n-1} \EE \bigg[ \sum_{i \neq k} \sum_{z}f(z,S_i)f(z,S_k) \hat \pv_\zv(z)^2 \bigg]\label{afvafgrsgh}\\ 
		&= \frac{n}{n-1} \sum_{i \neq k} \EE_{P_{S|Z} \times \hat \pv_\zv} \left[f(\tilde Z, S_i) \right]\EE_{P_{S|Z} \times \hat \pv_\zv} \left[f(\tilde Z, S_k) \right] \notag\\
		&~~- \frac{1}{n-1} \sum_{i \neq k} \EE_{\hat \pv_\zv} \left[ \EE_{P_{S|Z}}[f(\tilde Z,S_i)|\tilde Z ]\EE_{P_{S|Z}}[f(\tilde Z,S_k)|\tilde Z]\right] \label{guebaywrabwr}\\ 
		&= 2\binom{n}{2} \bigg(\frac{n}{n-1} \EE_{P_{S|Z} \times \hat \pv_\zv} \left[f(\tilde Z, S) \right]^2 \notag\\
		&~~~~~~~~~~~~~~~~~ - \frac{1}{n-1} \EE_{\hat \pv_\zv} \left[ \EE_{P_{S|Z}} [f(\tilde Z,S)|\tilde Z]^2\right]\bigg),\label{secondterm}
	\end{align}
	where \eqref{gusuerat} follows from expanding the expectation when the type of the sequence is known and $P_{Z_iS_i}$ does not depend on index $i$. Observe that \eqref{gusuerat} is divided into two parts because it addresses $z_1,z_2$ being equal or not in the expression $f(z_1,S_i)f(z_2,S_k)$. Observe that there are two terms separating all cases depending on whether $z_1,z_2$ are equal or not. When they are not equal, the number of such possibilities is $n\hat \pv_\zv(z_1)n\hat \pv_\zv(z_2)$ while the number of choices is $n(n-1)$, yielding a probability equal to $\frac{n}{n-1}\hat \pv_\zv(z_1)\hat \pv_\zv(z_2)$. Similarly, when $z_1=z_2 = z $, the number of such possibilities is $n\hat \pv_\zv(z_1)(n\hat \pv_\zv(z_2)-1)$, while the number of choices remains $n(n-1)$, yielding a probability equal to $\frac{1}{n-1}\hat \pv_\zv(z_1)(n\hat \pv_\zv(z_2)-1)$.
Eq. \eqref{afvafgrsgh} follows by rearranging the terms. Additionally, \eqref{guebaywrabwr} follows by taking the expectation inside using Lemma \ref{bfehdwbae}. 
	Combining \eqref{firstterm} and \eqref{secondterm} with \eqref{ptgfaefqw} we get the result.
\end{IEEEproof}

\begin{corollary}\label{corollary1}
	Under the assumptions of Lemma \ref{bfehdwbae} we have
	\begin{align}
		\Var \bigg[\sum_{i=1}^{n} f(Z_i,S_i) \Big| \hat \pv_\zv\bigg] = n \EE_{\hat \pv_\zv}\big[\Var_{P_{S|Z}}[f(\tilde Z,S)|\tilde Z] \big],
	\end{align}
	where $\tilde Z$ is a random variable with distribution $\hat \pv_\zv $.
\end{corollary}
\begin{IEEEproof}
	\begin{align} \label{advfgafgsge}
		\Var & \bigg[\sum_{i=1}^{n} f(Z_i,S_i) \Big| \hat \pv_\zv\bigg] \notag\\
		&= \EE \bigg[\bigg(\sum_{i=1}^{n} f(Z_i,S_i)\bigg)^2 \Big| \hat \pv_\zv \bigg] - \EE \bigg[\sum_{i=1}^{n} f(Z_i,S_i) \Big| \hat \pv_\zv \bigg]^2 \\ \label{ffadvtager}
		&= n^2 \EE_{P_{S|Z} \times \hat \pv_\zv}[f(\tilde Z,S)] ^2 +  n \EE_{P_{S|Z} \times \hat \pv_\zv}[f(\tilde Z,S)^2]\notag\\
		&~~ - n\EE_{\hat \pv_\zv} \left[ \EE_{P_{S|Z}} [f(\tilde Z,S)|\tilde Z]^2\right] -  n^2 \EE_{P_{S|Z} \times \hat \pv_\zv}[f(\tilde Z,S)] ^2\\
		&= n  \EE_{P_{S|Z} \times \hat \pv_\zv}[f(\tilde Z,S)^2] - n\EE_{\hat \pv_\zv} \left[ \EE_{P_{S|Z}} [f(\tilde Z,S)|\tilde Z ]^2\right] \\
		& = n \EE_{\hat \pv_\zv}[\Var_{P_{S|Z}}[f(\tilde Z,S)|\tilde Z] ]
	\end{align}
	where \eqref{advfgafgsge} follows from the  definition of variance, and \eqref{ffadvtager} follows by directly using Lemmas \ref{bfehdwbae} and \ref{lemma2}. 
\end{IEEEproof}

\begin{lemma}\label{setlemma}
	Let $(Z_i,S_i), i = 1,2,\dotsc, n$ be i.i.d random variables, $\zv = (Z_1,Z_2,\dotsc,Z_n)$ and $\Ac \subset \Pc^n_{\Zc}$ then 
	\begin{align} \label{muvar}
		&\EE \bigg[\sum_{i=1}^{n} f(Z_i,S_i) \Big |\Ac \bigg] \geq n \min_{\hat \pv_\zv \in \Ac} \EE_{P_{S|Z} \times \hat \pv_\zv} \left[ f(\tilde Z,S)\right]
	\end{align}
\end{lemma}
\begin{IEEEproof}
	We have
	\begin{align} \label{awdregvfrg}
		&\EE\bigg[\sum_{i=1}^{n} f(Z_i,S_i) \Big |\Ac\bigg] \notag\\
		&= \frac{1}{\PP(\Ac)} \EE\bigg[ \bigg(\sum_{i=1}^{n} f(Z_i,S_i)\bigg) \indicator \{\hat \pv_\zv\in \Ac \} \bigg] \\ 
		& \geq   \min_{\hat \pv_\zv\in \Ac} \EE\bigg[ \bigg(\sum_{i=1}^{n} f(Z_i,S_i)\bigg) \Big|\hat \pv_\zv \bigg]
		\\ \label{dracdvvh}
		& = n \min_{\hat \pv_\zv \in \Ac} \EE_{P_{S|Z} \times \hat \pv_\zv} \left[ f(\tilde Z,S)\right],
	\end{align}
	where \eqref{awdregvfrg} follows from the definition of conditional expectation and  \eqref{dracdvvh} follows from using Lemma \ref{bfehdwbae}.
\end{IEEEproof}

The next result introduces a lower bound on the variance.

\begin{lemma}
With the above assumptions we have
\begin{align}
\Var \bigg[\sum_{i=1}^{n} f(Z_i,S_i) \Big |\Ac\bigg] &\geq \min_{\tilde \pv_Z \in \Ac} \Var \bigg[\sum_{i=1}^{n} f(Z_i,S_i) \Big |\tilde \pv_Z \bigg] \\
&= n \min_{\tilde P_Z \in \Ac} \EE_{\tilde P_Z}\big[\Var_{P_{S|Z}}[f(\tilde Z,S)|\tilde Z] \big]
\end{align}
\end{lemma}

\begin{IEEEproof}
To show this we use the law of total variance which is stated below as a reminder. For any two random variable $X,Y$ we have 
\begin{align}
\Var[X] = \EE_{Y}[\Var[X|Y]] + \Var_{Y}[\EE[X|Y]]
\end{align}

As a result, by setting $Y = \hat \pv_Z$ meaning the random variable that denoted the type of the random variable $Z$. Then, we have

\begin{align}
&\Var \bigg[\sum_{i=1}^{n} f(Z_i,S_i) \Big |\Ac\bigg] \notag\\
&= \EE_{Y}\bigg[\Var \bigg[\sum_{i=1}^{n} f(Z_i,S_i) \Big |\Ac, Y \bigg]\bigg] \notag\\
&~~~~~~~~~~~~~~~+ \Var_Y\bigg[\EE \bigg[\sum_{i=1}^{n} f(Z_i,S_i) \Big |\Ac, Y \bigg]\bigg] \\
& \geq \EE_{Y}\bigg[\Var \bigg[\sum_{i=1}^{n} f(Z_i,S_i) \Big |\Ac, Y \bigg]\bigg] \\
&\geq \min_{\tilde \pv_Z \in \Ac} \Var \bigg[\sum_{i=1}^{n} f(Z_i,S_i) \Big |\tilde \pv_Z \bigg] \\
& = n \min_{\tilde P_Z \in \Ac} \EE_{\tilde P_Z}\big[\Var_{P_{S|Z}}[f(\tilde Z,S)|\tilde Z] \big].
\end{align}
\end{IEEEproof} 
In the next lemma we prove a concentration inequality for the same setting. In particular, we show that by conditioning on the type of a sequence we get a sub-Gaussian random variable. We prove the tail bound for the sum $\sum_{i=1}^{n} f(Z_i,S_i)$ when conditioned on the type $\hat \pv_\zv$.\\
The following lemma is an application of the Hoeffding's lemma but we prove it for completeness.
\begin{lemma} \label{avgfemfaokeaa}
	Let $f,Z_i,S_i$ be defined the same as Lemma \ref{bfehdwbae}. Further assume for all $z$ we have $a \leq f(z,S) - \EE[f(z,S)] \leq b$. Then,
	\begin{align} \label{cejofejfoefoo}
		\PP\bigg[\Big|\sum_{i=1}^{n} f(Z_i,S_i) - \mu\Big| \geq \xi \,\Big| \hat \pv_\zv\bigg] \leq 2e^{\frac{-\xi^2}{n(b-a)^2} },
	\end{align}
\end{lemma}
where $\mu = \EE\big[\sum_{i=1}^{n} f(Z_i,S_i) | \hat \pv_\zv \big]$.
\begin{IEEEproof}
Assume $\mu_\zv = \EE[f(z,S)] $, then for any $\lambda>0$
	\begin{align} \label{fefoejdfeoid}
		\PP\bigg[\sum_{i=1}^{n} f(Z_i,S_i) &- \mu \geq \xi \Big| \hat \pv_\zv\bigg] \notag\\
		&= \PP\left[e^{\lambda \left( \sum_{i=1}^{n} f(Z_i,S_i) - \mu \right) } \geq e^{\lambda\xi} \Big| \hat \pv_\zv\right] \\ \label{xvasufdeuf}
		&\leq  \frac{\EE\left[e^{\lambda \left( \sum_{i=1}^{n} f(Z_i,S_i) - \mu \right)} \Big| \hat \pv_\zv\right] }{e^{\lambda \xi}} \\ \label{adfjeuafnef}
		& = \frac{\prod_{z} \EE\left[e^{\lambda(f(z,S) - \mu_z)} \right]^{n \hat \pv_\zv(z)} }{e^{\lambda \xi}} \\
		&\leq  \frac{\prod_{z} e^{\frac{1}{8}\lambda^2(a- b)^2 n\hat \pv_\zv(z)}  }{e^{\lambda \xi}} \\ \label{bcaofdekfoe}
		& =  \frac{ e^{\frac{1}{8}\lambda^2(a- b)^2}  }{e^{\lambda \xi}}  \\ \label{femdoefbeudkpw}
		& \leq   e^{\frac{-\xi^2}{n(b-a)^2} },
	\end{align}
	where \eqref{xvasufdeuf} is derived for Markov's inequality. Additionally, \eqref{adfjeuafnef} follows from noticing that frequency of $f(z,S_i)$ appearing in the expression $\sum_{i=1}^{n} f(Z_i,S_i)$ for some $i$ is exactly $n \hat \pv_\zv(z)$ and because $S_i$s are i.i.d. the index of the appearance does not impact the moment generating function. Moreover, \eqref{bcaofdekfoe} is followed from setting $\lambda = \frac{4\xi}{n(b-a)^2}$. Therefore, \eqref{cejofejfoefoo} follows from
	\begin{align}
	&\PP\bigg[\Big|\sum_{i=1}^{n} f(Z_i,S_i) - \mu\Big| \geq \xi \,\Big| \hat \pv_\zv\bigg] \notag\\
	&~~~~~~~~~~~~~~~ = \PP\bigg[\sum_{i=1}^{n} f(Z_i,S_i) - \mu \geq \xi \Big| \hat \pv_\zv\bigg] \notag\\
	&~~~~~~~~~~~~~~~~~~~~~+ \PP\bigg[\sum_{i=1}^{n} f(Z_i,S_i) - \mu \leq -\xi \Big| \hat \pv_\zv\bigg]
	\end{align}
	and writing the same steps \eqref{fefoejdfeoid}--\eqref{femdoefbeudkpw} for $\PP\bigg[\sum_{i=1}^{n} f(Z_i,S_i) - \mu \leq -\xi \Big| \hat \pv_\zv\bigg]$.
\end{IEEEproof}

\bibliographystyle{IEEEtran}
\bibliography{bibliography}


\begin{IEEEbiographynophoto}{Ehsan Asadi Kangarshahi}
		 received both Bachelor Degrees in Mathematics and Electrical Engineering from Sharif University of Technology in 2017 and the Ph.D. from the University of Cambridge in 2022.
		
		From Oct.--Dec. 2017 he was a research intern at \'Ecole Polytechnique F\'ed\'erale de
		Lausanne (EPFL). From Jan. 2018- Oct. 2021 he was a Ph.D. student at the Department
		of Engineering, University of Cambridge, where he was also a member of Trinity Hall.		
		His research interests are in the areas of
		information theory, communication theory and statistics.
		
	\end{IEEEbiographynophoto}

\begin{IEEEbiographynophoto}{Albert Guill\'en i F\`abregas}
(S--01, M--05, SM--09, F--22) received the Telecommunications Engineering Degree and
the Electronics Engineering Degree from Universitat Polit\`ecnica de
Catalunya and Politecnico di Torino, respectively in 1999, and the Ph.D.
in Communication Systems from \'Ecole Polytechnique F\'ed\'erale de
Lausanne (EPFL) in 2004.

In 2020, he returned to a full-time faculty position at the Department
of Engineering, University of Cambridge, where he had been a full-time
faculty and Fellow of Trinity Hall from 2007 to 2012. Since 2011 he has
been an ICREA Research Professor at Universitat Pompeu Fabra (currently
on leave). He has held appointments at the New Jersey Institute of
Technology, Telecom Italia, European Space Agency (ESA), Institut
Eur\'ecom, University of South Australia, Universitat Pompeu Fabra, University of Cambridge, as
well as visiting appointments at EPFL, \'Ecole Nationale des
T\'el\'ecommunications (Paris), Universitat Pompeu Fabra, University of
South Australia, Centrum Wiskunde \& Informatica and Texas A\&M
University in Qatar. His specific research interests are in the areas of
information theory, communication theory, coding theory, statistical inference.

Dr. Guill\'en i F\`abregas is a Fellow of the Institute for Mathematics and its Applications (IMA) and a Member of the Young Academy of Europe. He received Starting and Consolidator Grants from the European
Research Council, the Young Authors Award of the 2004 European Signal
Processing Conference, the 2004 Best Doctoral Thesis Award from the
Spanish Institution of Telecommunications Engineers, and a Research
Fellowship of the Spanish Government to join ESA. Since 2013 he has been an Editor of the
Foundations and Trends in Communications and Information Theory, Now
Publishers and was an Associate Editor of the \sc{IEEE Transactions on
Information Theory} (2013--2020) and \sc{IEEE Transactions on Wireless
Communications} (2007--2011).
\end{IEEEbiographynophoto}

\end{document}